Title: Identifying Therapeutic Targets for Triple-Negative Breast Cancer using a Novel Mechanistic Mathematical Model of the Tumor Microenvironment

Authors: Kyle Adams,[1,5] Julia Bruner,[2,5] Salma Ameziane,[1,5] Ashley Brown,[3] Mohammed Gbadamosi,[4] Helen Moore[5]

1. Department of Mathematics, College of Liberal Arts and Sciences, University of Florida, Gainesville, FL, USA
2. Department of Surgery, College of Medicine, University of Florida, Gainesville, FL, USA
3. Institute for Therapeutic Innovation, Department of Medicine, College of Medicine, University of Florida, Gainesville, FL, USA
4. Department of Pharmacotherapy and Translational Research, College of Pharmacy, University of Florida, Gainesville, FL, USA
5. Laboratory for Systems Medicine, Department of Medicine, College of Medicine, University of Florida, Gainesville, FL, USA

Corresponding authors: Kyle Adams and Helen Moore

## Abstract

Triple-negative breast cancer (TNBC) is an aggressive disease with high mortality and limited treatment options, due to its lack of receptors that have targeted therapies available. The tumor microenvironment (TME) plays a critical role in TNBC progression and therapeutic resistance. In this work, we developed a novel mathematical model to describe key cellular interactions within the TNBC TME, informed by current literature and expert input. Our model consists of a system of ordinary differential equations representing five interacting cell populations: M2 macrophages, cancer-associated fibroblasts, TNBC tumor cells, cytotoxic T lymphocytes, and regulatory T cells. We performed global sensitivity analysis to determine which model parameters most-strongly influence tumor burden over a clinically-relevant treatment timeframe. The pathways associated with the most-influential parameters correspond to biological mechanisms that are consistent with known and emerging therapeutic strategies in TNBC, including stromal-mediated tumor support. These results highlight key regulatory interactions within the TNBC TME and provide a quantitative framework for hypothesis generation and future investigation of combination treatment strategies.

## Introduction

In 157 countries out of the 185 studied by the Global Cancer Observatory, breast cancer was the most-commonly diagnosed cancer in women in 2022.[1] The American Cancer Society estimates that a woman in the United States has approximately a 1 in 8 chance of developing breast cancer in her lifetime.[2] This disease affects women of all races and ethnicities, and is the second-most deadly cancer in women after lung cancer.[3] The importance of breast cancer research is hard to overstate.

In this work, we focus on triple-negative breast cancer (TNBC) due to its aggressiveness, its low survival rates, and the fact that it is the most immunogenic of the breast cancer subtypes.[4] Approximately 10-15% of all breast cancers are triple-negative, which refers to cancer cells that lack three key receptors:

estrogen receptor (ER), progesterone receptor (PR), and HER2 receptor.[5] Depending on the cancer stage at diagnosis, chemotherapy, immunotherapy, radiation, and surgery are current treatment options for people with TNBC.[5] However, due to negativity for ER, PR, and HER2, TNBC tumors are not responsive to therapies that are available to target these receptors, and so there are fewer treatment options compared to other breast cancer subtypes.[5]

The TNBC tumor microenvironment (TME) has been an area of recent focus.[4,6] Our goal in this work was to contribute to this research by developing a mechanistic math model describing specific components of the TME of TNBC. Different types of mathematical models have been widely used in medicine with goals such as predicting the hospitalization needs of COVID-19 patients,[7] optimizing treatment schedules for cancer patients,[8,9] and exploring treatment resistance.[10] Recently, immunotherapy has shown significant potential as a treatment for TNBC,[11] which underscores the importance of including tumor-immune dynamics and focusing on the TME.

In this work, we provide a comprehensive literature review of dozens of existing relevant cellular-level mathematical models of breast cancer, and summarize their key components in a table. This provides additional evidence of elements important to include in the cellular dynamics of breast cancer, as well as gaps in the models, analyses, and interpretations. We then share the foundations and justifications for the novel model we developed, and describe the specific cell populations included and their interactions in the TNBC TME. We provide detailed justifications for model parameter estimates, the majority of which are calculated from literature data or information. We also perform global sensitivity analysis and discuss the implications of the most-influential parameters identified for potential therapeutic targeting.

While we did not explicitly simulate therapeutic interventions, our goal was to identify biologically-influential pathways that regulate tumor burden over clinically-relevant timescales. The results of this work are intended to generate hypotheses regarding promising intervention strategies, which may be evaluated in future computational, experimental, or clinical studies.

## Mechanistic Mathematical Models Of Breast Cancer

Many mechanistic math models for breast cancer have been proposed previously, with different approaches and goals. These include ODE models, fractional models, stochastic models, agent-based models (ABMs), and partial differential equation (PDE) models. In this section, we briefly describe the cellular-level models of breast cancer that are most relevant to ours, and we summarize them in **Table 1.**

Roe-Dale et al. proposed an ODE model to explain clinical results for sequential vs. alternating regimens of doxorubicin and cyclophosphamide, methotrexate, and fluorouracil therapy.[12] To study the effects of estrogen on breast cancer, Mufudza et al. first developed and analyzed stability of an estrogen-free competition model of healthy breast epithelial cells, breast cancer cells, and immune cells; they reanalyzed their model after adding estrogen dynamics, suggesting that the addition of excess estrogen promotes tumor growth.[13] Abernathy et al. changed some of the equations and extended the model of Mufudza et al. by adding cancer stem cells.[14] The Abernathy et al. model was later adapted by Sukpol et al. to understand intratumoral heterogeneity and competition.[15] The Abernathy et al. model was also the foundation for a model by Solís-Pérez et al. that used fractional derivatives.[16] Jarrett et al. created an ODE model of HER2+ breast cancer and included effects of trastuzumab, a therapy that targets HER2+ cells they also included effects on immune cells, and fit their model to mouse data.[17] Wei developed their own ODE model for ER+ BC that included immune cells, and fit the model to mouse and human data,[18] and then used the model to simulate treatment with AZD9496 and palbociclib.[19] They later modified this model for TNBC, and explored treatment with an anti-PD-1 immune checkpoint inhibitor.[20] Wang et al. developed a large ODE model of disease-therapy dynamics, sometimes referred to as a quantitative systems pharmacology (QSP) model, with 343 equations over four compartments (central, peripheral, tumor, and tumor-draining lymph node) and simulated a clinical trial with anti-CTLA-4 and anti-PD-L1 immunotherapies;[21] this model was later expanded with a focus on TNBC.[22] The same group used this TNBC QSP model to study dynamics of tumor-associated macrophages, comparing virtual patients to clinical data.[23] They then added metastases to the model to look for predictive biomarkers of PD-1 inhibition.[24]

Padmanabhan et al. reviewed many mathematical models of BC and created a general summarizing model of HER2+ BC that included HER2 and PD-1/PD-L1 pathways.[25] Mehdizadeh et al. proposed an ODE model that includes proliferative cancer cells, quiescent cancer cells, CD8+ T cells, and natural killer cells to investigate immunological dormancy of TNBC, and compared model simulations to mouse data.[26] Mirzaei et al. developed an ODE model of tumor-immune dynamics to study the effect of immune cell infiltration, particularly macrophages, CTLs, helper T cells, Tregs, adipocytes, necrotic cells, and various cytokines.[27] The authors later fit their model to mouse data.[28] Davenport et al. developed a model of ODEs to describe the dynamics among TNBC tumor proliferation, necrosis, apoptosis, vasculature, and immune response to doxorubicin and paclitaxel, and fit the model to mouse tumor volume data.[29] Akman et al. developed an ODE model to investigate the relationship between diet and anti-hormonal therapy, which they fit to mouse tumor volume data; they then applied optimal control to determine best regimens.[30] Syme et al. developed a model of ODEs to describe the immune response of mouse CD4+ and CD8+ T cells to a TNBC tumor while being treated with anti-PD-1 and anti-CTLA-4 checkpoint inhibitors and fit the model to mouse data.[31] Akil et al. applied optimal control to their ODE model of ER+ breast cancer, using ketogenic diet, hormonal therapy, and immunotherapy as their controls.[32]

Based on a series of tumor models by Anderson,[33,34] Enderling et al. proposed a PDE model describing breast cancer tumor growth and invasion to investigate radiotherapy and local recurrence.[35] Knútsdóttir et al. proposed a PDE model to aid in drug design by modeling macrophages and a few of their cytokine secretions.[36] Weis et al. used a reaction-diffusion PDE model to predict the response of breast cancer patients to neoadjuvant therapy.[37] Miniere et al. proposed a reaction-diffusion PDE model to predict the response of breast cancer cells to doxorubicin treatment.[38] Christenson et al. developed a PDE model to optimize neoadjuvant chemotherapy regimens.[39] Finally, based on an earlier model,[40] Norton et al. developed ABMs to understand the relationship between expression of the biomarker CCR5, cancer stem cells, and hypoxia in TNBC,[41] and the effect of TNBC heterogeneity.[42]

Each of these models differs from ours in the cell populations that are included. In particular, none of the existing mathematical models we found included cancer-associated fibroblasts (CAFs), despite evidence of their importance in breast cancer dynamics.[43–47] Some of these models include a treatment, whereas the goal of our work was to establish a model of key TNBC TME dynamics for analysis of key drivers of tumor size, and to expand and further analyze it later.

| **Mathematical Models of Breast Cancer Cited in this Work** | | | | | |
|---|---|---|---|---|---|
| **Authors (Year)** | **Model Type** | **Cancer Type** | **Model Components** | **Notes** | **Reference** |
| Roe-Dale et al. (2010) | ODE | BC | Sensitive cancer cells at different life-cycle phases, resistant cancer cells at different life-cycle phases | Math model developed to explain clinical benefits of sequential over alternating chemotherapy | 12 |
| Mufudza et al. (2012) | ODE | Luminal BC | Healthy breast cells, breast cancer cells, immune cells (combined NK cells and CD8+ T cells), estrogen | Stable equilibria analysis; cancer cell behavior with and without estrogen | 13 |
| Abernathy et al. (2020) | ODE | BC | Cancer stem cells, other cancer cells, healthy cells, immune cells, estrogen | Tumor-immune competition and effects of estrogen (based on Mufudza at el. 2012 model) | 14 |
| Sukpol et al. (2024) | ODE | BC | Two competing BC cell types, each with cancer stem cells and other cancer cells, and NK cells | Analysis of competitive interactions, equilibria/stability analysis | 15 |
| Solís-Pérez et al. (2019) | Fractional | BC | Cancer stem cells, other cancer cells, healthy cells, immune cells, excess estrogen | Fractional model version of Abernathy et al. 2020 model | 16 |
| Jarrett et al. (2019) | ODE | HER2+ BC | Cancer cells, immune response, necrosis, vasculature, hypoxia | Model fit to mouse data, including effects of trastuzumab | 17 |
| Wei (2019) | ODE | ER+ BC | Cancer cells, NK cells, CTLs, white blood cells, estradiol | Model fit to mouse and human data | 18 |
| Wei (2020) | ODE | ER+ BC | Cancer cells, NK cells, CTLs, white blood cells, estradiol, AZD9496, palbociclib | Simulations of AZD9496 and palbociclib treatment dynamics (therapies were added to Wei 2019 model) | 19 |
| Wei (2023) | ODE | TNBC | Cancer cells, active NK cells, active CTLs, circulating lymphocytes, PD-L1 concentration, PD-1 concentration, anti-PD-L1 drug concentration | Immune checkpoint inhibition modelling | 20 |
| Wang et al. (2019) | QSP/ODE | ER+ and TNBC | 343 variables, including T cells and cancer cells with different immune checkpoint expression levels | Simulated a clinical trial with anti-CTLA-4 and anti-PD-L1 immunotherapies | 21 |
| Wang et al. (2021) | QSP/ODE | TNBC | Added atezolizumab and nab-paclitaxel to Wang et al. 2019 model | Simulations to predict efficacy of atezolizumab and nab- | 22 |

| | | | | paclitaxel | |
|---|---|---|---|---|---|
| Wang et al. (2022) | QSP/ODE | TNBC | Added macrophages to Wang et al. 2019 model | Compared a simulated virtual population to clinical data | 23 |
| Arulraj et al. (2023) | QSP/ODE | TNBC | Added metastases to Wang et al. 2019 model for a total of 641 equations | Identified predictive biomarkers for PD-1 inhibition | 24 |
| Padmanabhan et al. (2020) | ODE | HER2+ | Cancer cells, immune cells, vascular delivery, treatment | Studied crosstalk between HER2 and PD-1/PD-L1 | 25 |
| Mehdizadeh et al. (2021) | ODE | TNBC | Proliferative cancer cells, quiescent cancer cells, CD8+ T cells, NK cells | Simulated immunological dormancy, compared to mouse data | 26 |
| Mirzaei et al. (2021, 2022) | ODE | BC | Cancer cells, naive and activated macrophages, naive T cells, CD4+ T cells, CTLs, Tregs, naive and activated dendritic cells, cancer-associated adipocytes, necrotic cells, HMGB1, IL-12, IL-10, estrogen, IFN-gamma, IL-6 | Simulations of tumor–immune microenvironment dynamics; model was fit to mouse RNA-Seq data | 27,28 |
| Davenport et al. (2022) | ODE | TNBC | Cancer cells (proliferating, necrotic, and apoptotic), vasculature, immune cells, doxorubicin, paclitaxel | Studied tumor response to combination chemotherapy; fit model to mouse tumor volume data | 29 |
| Akman et al. (2024) | ODE | ER+ BC | Cancer cells (resistant and sensitive to estrogen deprivation), estrogen, fat | Models with diet and anti-hormonal therapy; fit to mouse tumor size data; optimal control treatments calculated | 30 |
| Syme et al. (2025) | ODE | TNBC | Cancer cells, APCs, CD8+ T cells, CD4+ T cells, anti-PD-1 and anti-CTLA-4 therapies | Model of tumor response to combination immunotherapy; fit to | 31 |

| | | | | mouse tumor volume and CD4+/CD8+ T cell counts | |
|---|---|---|---|---|---|
| Akil et al. (2025) | ODE | ER+ BC | Healthy breast cells, breast cancer cells, lymphocytes, estrogen, immunotherapy | Optimal control for diet, immunotherapy, & hormone therapy | 32 |
| Enderling et al. (2006) | PDE | BC | Cancer cells, extracellular matrix, matrix-degrading enzymes, radiotherapy | Tumor invasion and radiotherapy | 35 |
| Knútsdóttir et al. (2014) | PDE | BC | Cancer cells, macrophages, colony-stimulating factor-1, epidermal growth factor | Spatial tumor–immune interactions | 36 |
| Weis et al. (2015) | PDE | BC | Cancer cells | Reaction-diffusion model fit to human data, simulated with neoadjuvant therapy | 37 |
| Miniere et al. (2025) | PDE | ER+ BC | Cancer cells, doxorubicin | Reaction-diffusion model fit to in vitro data | 38 |
| Christenson et al. (2024) | PDE | TNBC | Cancer cells, doxorubicin, cyclophosphamide | Model fit to tumor size data; optimal control applied for neoadjuvant therapies | 39 |
| Norton et al. (2017) | ABM | TNBC | Cancer cells, hypoxia, CCR5, anti-CCR5 therapy | Model simulations of CCR5 and hypoxia-driven tumor evolution | 41 |
| Norton et al. (2018) | ABM | TNBC | Cancer stem cells, CCR5+/- cancer cells, hypoxia, macrophages, fibroblasts, angiogenesis | TNBC heterogeneity | 42 |

**Table 1.** Summary of relevant mechanistic mathematical models of breast cancer cell dynamics discussed in this paper. ABM = agent-based model, APC = antigen-presenting cell, BC = breast cancer, CTL = cytotoxic T lymphocyte, ECM = extracellular matrix, NK = natural killer, ODE = ordinary differential equation, PDE = partial differential equation, QSP = quantitative systems pharmacology, RNA-Seq = RNA sequencing, TNBC = triple-negative breast cancer.

# Model and Justifications

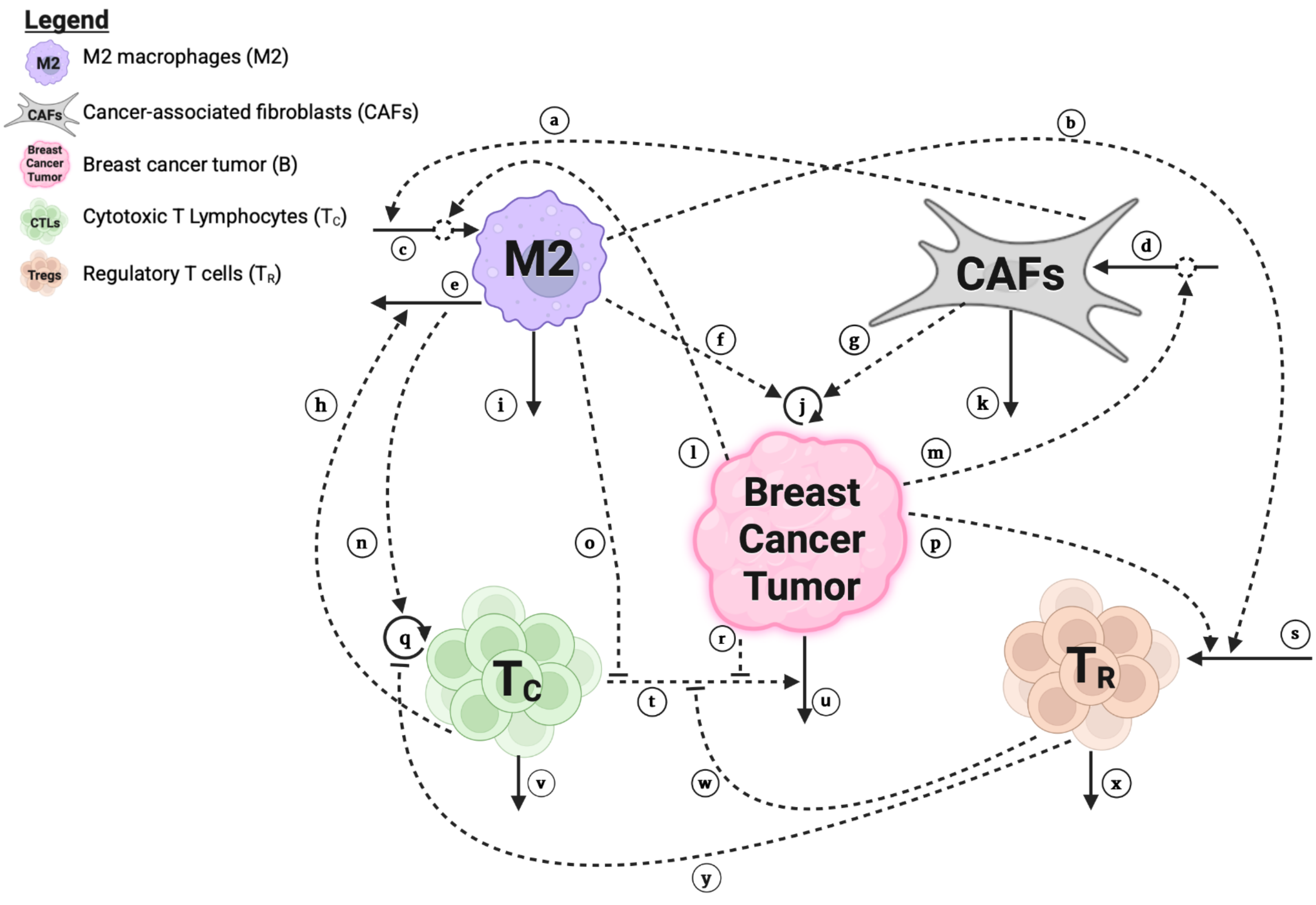

**Fig. 1. Model diagram with cells and interactions.** Each of the five cell populations, M2 macrophages (M2), cancer-associated fibroblasts (CAFs), triple-negative breast cancer tumor cells (Breast Cancer Tumor), activated cytotoxic T lymphocytes ($T_C$), and regulatory T cells ($T_R$), is represented by a shape and its abbreviation. Solid arrows represent a change in cell number. Solid arrows pointing away from a cell depict loss of that cell's population (due to natural death, or repolarization in the case of **arrow e**). Solid arrows pointing toward a cell population represent an increase in that population's number. This could be due to polarization, transformation (specifically normal fibroblasts transforming into CAFs in arrow **d**), recruitment, or proliferation in the case of a circular solid arrow. Dotted curves with arrows represent a boosting effect on the mechanism they are pointing to. Dotted curves with a flat end represent an inhibition effect on the mechanism they are pointing to. The dynamics and mechanisms are discussed in detail in the Cell Populations & Dynamics section. This figure was created with Biorender.com.

## Cell Populations & Dynamics

The TNBC tumor microenvironment is complex, with many types of cells contributing pro-tumor or anti-tumor effects. In this work, our goal was to establish a simple model capturing the dynamics of the most-important cell populations in the TNBC TME. **Fig. 1** shows the diagram for the model we built. Our model is small by design, and is representative of "net effects" that include effects from cells and mechanisms that are not explicitly represented in the model. We included cell populations if they were clearly key components in TNBC dynamics, if their importance had scientific consensus, and if their dynamics may be affected by current or hypothesized treatments. This section describes each cell type in the model, and processes (represented by arrows) that affect its cell population numbers. A description of each arrow is included in the cell type that it affects, in the same order that the arrow labels appear in Equations (1)-(5).

**M2 Macrophages (M2)**

Several studies have indicated the important roles that tumor-associated macrophages (TAMs) play in the TME of TNBC.[48–51] TAMs originate from bone marrow-derived monocytes which circulate in peripheral blood before being recruited to the TME by tumor cells or by host immune cells. In tissue, they differentiate into macrophages of different phenotypic classes, in a process called polarization. TAMs may refer to both M1-like macrophages, which have anti-tumor, pro-inflammatory activity, and M2-like macrophages, which have pro-tumor, anti-inflammatory activity. While there is a spectrum of macrophage phenotypes,[52] in our model, we labeled all pro-tumor macrophages as M2 macrophages. That is, M2 macrophages in our model could refer to either an M2 macrophage or an M2-like macrophage. We included M2 macrophages in our model as they make up a majority of TAM populations within tumors,[53] and the proportion of TAMs that are M2 macrophages has been reported to increase as the tumor progresses.[54] The source of M2 macrophages **(arrow c)** represents both direct polarization of undifferentiated monocytes into M2 macrophages as well as the repolarization of M1 macrophages to the M2 phenotype. In either case, the source of *tumor-associated* M2 macrophages **(arrow c)** is dependent on the presence of tumor cells **(arrow l)**. Polarization of TAMs toward the M2 phenotype is driven by anti-inflammatory cytokines and signals from other cell types including CAFs **(arrow a)** and TNBC tumor cells **(arrow l)**. Cohen et al.[43] showed that the protein Chi3L1 secreted by CAFs causes macrophages to migrate to the TME and polarizes them to be M2. Several studies also discuss the release of TGF-β from CAFs,[44,55] which has been shown to polarize M1 to M2 macrophages.[56,57] Breast cancer cells secrete IL-2, IL-10, TGF-β, and M-CSF, which polarize M1-like macrophages to M2 macrophages.[4,48] Also, TNBC specifically secretes more G-CSF compared to other breast cancer subtypes, which is also a promoter of M1 to M2 polarization.[58]

We chose to capture the dynamics of M1 macrophages indirectly by including the possible repolarization of M2 macrophages to M1 macrophages (**arrow e**). Repolarization of M2 macrophages to M1 macrophages has been shown to occur in osteosarcoma,[59] and with a treatment (matairesinol) in triple-negative breast cancer.[59,60] Yuan et al. found that higher infiltration of TAMs in TNBC is associated with a “significantly higher risk of distant metastasis”, and lower rates of overall survival (OS) and disease-free survival (DFS).[50] Another TME dynamic implicitly captured is CTLs boosting polarization of M2 macrophages to M1 macrophages **(arrow h)**. Because the TME is generally immunosuppressive, the rate of polarization of M2 to M1 macrophages in vivo is low. However, administration of immunotherapies leading to T cell activation and engagement with tumor cells leads to greater tumor infiltration with M1 macrophages.[61,62] While we represent this effect with **arrow h**, we note that this observed increase in macrophages is also likely to occur through recruitment of M1 macrophages**.** Modelling the TNBC TME with both M1 and M2 macrophages is useful as conversion of TAMs to M1-like macrophages is a potential new therapy in triple negative breast cancer.[63] The natural loss of M2 macrophages upon completion of a finite lifespan is represented by **arrow i**.

**Cancer-Associated Fibroblasts (CAFs)**

CAFs are resident fibroblasts that are activated by tumor cells toward a pro-tumor state **(arrow d)**.[45,64] CAFs are highly heterogeneous.[64] For the purpose of our model, we consider any fibroblast that exhibits pro-tumor effects to qualify as a CAF. Because the physical and local occupation of CAFs within the tumor is responsible for structural changes such as remodeling the extracellular matrix and increasing the stiffness of the tumor,[65] we chose to model their total volume within the tumor rather than their density, as opposed to the immune cells in our model. The transformation from normal fibroblasts to CAFs is induced by tumor-secreted factors including TGF-β, PDGF, and CXCL12 **(arrow m, d)**.[46,47] The natural loss of CAFs upon completion of a finite lifespan is represented by **arrow k**.

Because of these effects, there has been growing interest in the role CAFs play in the tumor microenvironment of TNBC.[66–68] CAFs are found in several cancers,[64] and in TNBC there is a correlation between CAFs and tumor aggressiveness and disease recurrence.[69] CAFs are also implicated in macrophage dynamics, immune suppression, and immune escape.[66] In some cases CAFs have been shown to induce resistance to treatments including radiotherapy, hormonal therapy, and chemotherapy.[68] Since TNBC often exhibits chemotherapeutic resistance and has an immunosuppressive tumor microenvironment, CAFs have been investigated recently as a therapeutic target to address these challenges.[70]

**Breast Cancer Cells (B)**

The breast cancer tumor size is modeled by logistic growth (**arrow j**), as is common for tumors,[71] since there should be some (possibly large, never attained) physical limit to the size of the tumor. Breast cancer tumor growth is boosted by M2 macrophages **(arrow f)** and CAFs **(arrow g)**. TAMs, which tend to be M2 macrophages, especially as the tumor progresses,[54] secrete growth factors including BFGF-2, EGF, PDGF, and TGF-β, which are responsible for tumor cell proliferation **(arrow f)**.[4] In the TME, CAFs' secretion of TGF-β and CXCL12 increases proliferation of breast cancer cells leading to enhanced tumor growth **(arrow g)**.[47]

We assume tumor cells have a finite lifespan **(arrow u)**.[72] CTLs kill tumor cells **(arrow t)** using the perforin-granzyme pathway and through expression of death ligands, including Fas ligand (FasL) and TNF-related apoptosis inducing ligand (TRAIL).[73] However, CTL cytotoxic activity is suppressed by several cells in the immunosuppressive TME. Multiple immune checkpoint ligands that M2 macrophages express, including PDL1 and B7-H4, have been implicated in blocking CTL activity **(arrow o)**.[74–76] M2 macrophages also produce TGF-β, which inhibits CTL function **(arrow o)**.[77] M2 macrophages can also inhibit CTL proliferation,[78–82] including through a mechanism (arginine metabolism) being explored as a therapy,[83] but to simplify the model and to reduce potential structural identifiability issues, we include this effect in our estimate for parameters involved in **arrow o**. Expression of the immune checkpoint ligand PD-L1 on TNBC cells can lead to exhaustion/inactivation of CTLs using PD-L1 **(arrow r)**.[84] Sawant et al. showed that subsets of Treg populations in the TME can express IL-10 and IL-35, leading to exhaustion of CD8+ tumor-infiltrating lymphocytes (**arrow w**).[85]

**Cytotoxic T lymphocytes (CTLs)**

Cytotoxic T lymphocytes, also known as CD8+ T cells, are a subset of tumor-infiltrating lymphocytes (TILs) responsible for direct immune-mediated killing of damaged, infected, or cancerous host cells. CTLs are critical players in the TME; they are a valuable prognostic metric and vital to anti-PD-1 therapy.[86,87] In this model, we specifically chose to model activated CD8+ T cells, that is, CD8+ T cells that exhibit functional cytotoxic activity against tumor cells. Upon recognition of tumor antigens, and with costimulatory signals from other immune cells, CTLs become activated, proliferate **(arrow q)**, and target cancer cells for death. Although M1 macrophages and CD4+ T cells are not explicitly included as cell populations in our model, their downstream effects in CTL activation are captured by **arrow n.** M1 macrophages are classically-activated antigen-presenting cells; once M2 macrophages are polarized to M1 macrophages (**arrow e**), they participate in T cell activation and clonal expansion (**arrow n**).[88] Breast cancer cells can upregulate PD-L1 expression, resulting in binding to PD-1 receptors on CTLs, inactivating the CTLs. It is estimated that 30% to 50% of all breast cancers have upregulated PD-L1.[89] Competition for IL-2 between Tregs and CTLs has been shown to directly inhibit CTL proliferation (**arrow y**).[90] Tay et al. reviewed Tregs as immunotherapy targets in cancer, supported by Tregs' potential to outcompete conventional T-cells for IL-2.[91] **Arrow v** represents the loss of CTLs, as they have a finite lifespan.[92]

**Regulatory T Cells (Tregs)**

Tregs, or CD4+CD25+FOXP3+ T cells, play an integral role in tumor progression.[78,93,94] Labeling studies by Vukmanovic-Sejic et al. in healthy volunteers supported the hypothesis that Tregs are limited in their self-renewal, and are likely continuously recruited from a pool of CD4+ memory T cells.[95] Thus we assume that the Tregs in our model are recruited to the TME from a source outside of our model. Tregs are generally recruited to the TME via chemokines (**arrow s**).[96] M2 macrophages also recruit Tregs to the TME **(arrow b)**. M2 macrophages secrete anti-inflammatory cytokines and chemokines, including TGF-β, which induce differentiation of naive helper T cells into Tregs.[97] In ovarian cancer studies, macrophage-derived chemokine CCL22 results in the recruitment of peripheral Tregs.[98] In claudin-low breast cancers, which are low in tight junction and adhesion proteins, and are generally found in patients with TNBC, Tregs are recruited through the chemokine CXCL12 generated by the tumor **(arrow p)**.[99] Chang et al. found that colon cancer cells express CCL5 which recruits Tregs through CCR5,[100] a mechanism further evaluated in breast cancer.[101] Tregs are also recruited by cytokines that M2 macrophages release.[97] **Arrow x** represents the loss of Tregs, as they have a finite lifespan.[92]

## The Model

$$(1)\frac{dM_2}{dt} = \overbrace{(\frac{\alpha_{B2}B}{\beta_{B2}+B})}^{\text{c,l}}\overbrace{(1+\frac{\alpha_{F2}F}{\beta_{F2}+F})}^{\text{a}} - \overbrace{\gamma_2 M_2}^{\text{e}}\overbrace{(1+\frac{\alpha_{C2}T_C}{\beta_{C2}+T_C})}^{\text{h}} - \overbrace{\delta_2 M_2}^{\text{i}}$$

$$(2)\frac{dF}{dt} = \overbrace{\sigma_F B(1-\frac{F}{K_F})}^{\text{d,m}} - \overbrace{\delta_F F}^{\text{k}}$$

$$(3)\frac{dB}{dt} = \overbrace{p_B B(1-\frac{B}{K_B})}^{\text{j}}\overbrace{(1+\frac{\alpha_{2B}M_2}{\beta_{2B}+M_2})}^{\text{f}}\overbrace{(1+\frac{\alpha_{FB}F}{\beta_{FB}+F})}^{\text{g}} - \overbrace{\delta_B B}^{\text{u}}\overbrace{(1+\frac{\alpha_{CB}T_C}{\beta_{CB}+T_C}}^{\text{t}}[\overbrace{(1-\frac{\alpha_{2CB}M_2}{\beta_{2CB}+M_2})}^{\text{o}}\overbrace{(1-\frac{\alpha_{BCB}B}{\beta_{BCB}+B})}^{\text{r}}\overbrace{(1-\frac{\alpha_{RCB}T_R}{\beta_{RCB}+T_R})}^{\text{w}}])$$

$$(4)\frac{dT_C}{dt} = \overbrace{p_C T_C(1-\frac{T_C}{K_C})}^{\text{q}}\overbrace{(1+\frac{\alpha_{2C}M_2}{\beta_{2C}+M_2})}^{\text{n}}\overbrace{(1-\frac{\alpha_{RC}T_R}{\beta_{RC}+T_R})}^{\text{y}} - \overbrace{\delta_C T_C}^{\text{v}}$$

$$(5)\frac{dT_R}{dt} = \overbrace{\sigma_R}^{\text{s}}\overbrace{(1+\frac{\alpha_{2R}M_2}{\beta_{2R}+M_2})}^{\text{b}}\overbrace{(1+\frac{\alpha_{BR}B}{\beta_{BR}+B})}^{\text{p}} - \overbrace{\delta_R T_R}^{\text{x}}$$

### Model Assumptions

Our model of the TNBC TME consists of a system of five ODEs, one for each cell population. A large majority of TNBC patients are diagnosed at Stages I, II, or III of the disease; the most-common stage at diagnosis is Stage II.[102] We have chosen to model TNBC beginning at Stage II, which is reflected in certain choices of the model diagram and equations discussed below. Because we plan to use our model to optimize combination drug regimens for TNBC in the future, our model reflects dynamics within the TME from the time of a Stage II diagnosis to 168 days after. 168 days reflects current treatment protocols lasting 24 weeks for TNBC.[103] Each equation in the model begins with an input/source of a cell population and ends with an output/loss of that population, where each source or loss may be multiplied by one or several interactions. M2 macrophages have two loss terms, natural loss and polarization to M1 macrophages. For biological plausibility, we assume that every rate in our model has a finite bound. For example, the boosting effect of CAFs polarizing M1 macrophages to M2 macrophages (**arrow a**) must be limited; that is, it is not biologically plausible for CAFs to boost polarization without bound. This assumption is reflected in the boosting and inhibiting mechanisms, which we model with Michaelis-Menten terms to ensure that all rates remain bounded. We also assume that all cells in our model have finite lifespans.

## Model Size

The TME is made up of numerous interacting cell types and signaling pathways, many of which have overlapping effects. For computational tractability, and to reduce structural identifiability challenges due to limited available data, we combined multiple mechanisms into single pathways when possible. For example, M2 macrophage-mediated suppression of cytotoxic T lymphocytes occurs through multiple mechanisms, including immune checkpoint signaling, cytokine release, and metabolic competition. Rather than modeling each mechanism explicitly, we represented their combined net effect through a single inhibitory pathway (**arrow o**). Similarly, the influence of M1 macrophages on T-cell activation is represented indirectly through macrophage polarization dynamics (**arrow n**) rather than as a separate M1 macrophage cell population.

These modeling choices reflect a balance between biological detail and parameter identifiability. Rather than representing individual mechanisms, our parameters represent "net effects", as mentioned in the "Cell Population & Dynamics" section. Our model size allows for robust analysis of parameter-level influence on tumor dynamics. Future work incorporating additional information and data may allow further mechanistic resolution of these aggregated effects.

## Equations

Here we describe in detail the equations we chose to represent the tumor-immune interactions in our model. The model mechanisms and references are also gathered in **Table 2**, for convenience.

**Equation (1):** **Arrow c** represents the source of M2 macrophages recruited to the TME. This includes M1 macrophages that are polarized to M2 macrophages,[104] and undifferentiated monocytes that are directly polarized to M2 macrophages, which regularly occurs in cancer.[104,105] We use **arrow l** to model that M2 macrophage recruitment is dependent on the existence of the TNBC tumor. Since we are modeling the TME of TNBC, we require that TNBC cells exist, in order to measure immune cells on or surrounding the tumor. We label the source of M2 macrophages as **arrow c,l** to reflect this dependency and we use a Michaelis-Menten term to model this process. We did not include a carrying capacity for M2 macrophages, as monocytes are continuously produced from the bone marrow. Instead, the M2 macrophage population is inherently bounded by the TNBC tumor's carrying capacity. This source is multiplied by **arrow a**, which represents the boost in the M2 macrophage population from CAFs polarizing M1 macrophages to M2 macrophages. We model the possible repolarization of M2 macrophages to M1 macrophages using **arrow e**, which is multiplied by **arrow h**, due to the boost that CTLs can have on this repolarization process.[106,107] This is followed by the natural loss of M2 macrophages, with the rate constant $\delta_2$ multiplied by $M_2$ to represent that M2 macrophages are lost in proportion to their population size.

**Equation (2):** CAFs are the most-abundant stromal cell type in the TNBC microenvironment.[108] Tumor cells influence normal fibroblasts to transform into CAFs,[109] represented by **arrow d,m**. Since CAFs are resident fibroblasts that are activated, rather than recruited to the environment, we include a carrying capacity term, $(1 - \frac{F}{K_F})$, to limit their growth. The other factors in this growth term represent a proportion of the TNBC cell level. **Arrow k** represents the death rate of CAFs.

**Equation (3):** The first term begins with proliferation of the breast cancer cells, represented by **arrow j**, which we assume can be modeled by logistic growth.[71] This is multiplied by the boosting effects from M2 macrophages **(arrow f)** and CAFs **(arrow g)**.[110] The loss term is multiplied by a boosting effect from CTLs directly killing tumor cells (**arrow t**).[73] There are multiple mechanisms that inhibit this CTL killing activity. We multiply by factors that represent M2 macrophages suppressing CTL activity **(arrow o)**,[51] TNBC cells inactivating CTLs using PD-L1 **(arrow r)**,[84] and Tregs suppressing CTLs **(arrow w)**.[94]

**Equation (4):** The first term begins with proliferation of activated CTLs, term **q**, which we assume can be modeled by logistic growth because they divide rapidly (but not without bound) via clonal expansion.[111] When M2 macrophages are repolarized into M1 macrophages, the resulting M1 macrophages will resume antigen presentation, resulting in CTL proliferation.[112] Since M1 macrophages are not explicitly represented in the model, we model this boost by multiplying $\alpha_{2C}$ by $\gamma_2 M_2$, the repolarization rate of M2 to M1. Therefore, we can write term **n** as

$$(1+\frac{\alpha_{2C}\gamma_2 M_2}{\beta_{2C}+\gamma_2 M_2})(1+\frac{\alpha_{2C}\gamma_2 M_2}{\beta_{2C}+\gamma_2 M_2})$$

However, since $\gamma_2$ is a constant, we can divide the fraction by $\gamma_2$ and obtain:

$$(1+\frac{\alpha_{2C} M_2}{\frac{\beta_{2C}}{\gamma_2}+M_2})(1+\frac{\alpha_{2C} M_2}{\frac{\beta_{2C}}{\gamma_2}+M_2})$$

Since the fraction $\frac{\beta_{2C}}{\gamma_2}$ is still a constant, we renamed it as $\beta_{2C}$ for simplicity. Although M2 macrophages inhibit CTL proliferation,[113] we did not include this in the model, as it would cause an identifiability issue with **arrow o**. Instead, we use **arrow o** as a catch-all term for any inhibition M2 macrophages have on CTLs. The last term in this equation is for the natural loss of activated CTLs, which represents the net effect of both CTL death and inactivation **(arrow v)**.

**Equation (5):** The **s** term represents the recruitment of Tregs. Terms **b** and **p** are the boosting effects that M2 macrophages and TNBC cells, respectively, have on Treg recruitment.[54,114] This is followed by a natural loss term (**arrow x**).

| Model Arrows | | |
|---|---|---|
| **Label** | **Description** | **References** |
| a | CAFs boost differentiation of monocytes to M2 macrophages and polarization of M1 to M2 | [4,43,44,55–57] |
| b | M2 macrophages boost recruitment of Tregs | [97,98] |
| c | M0 macrophages differentiate into M2 macrophages, and M1 macrophages polarize to M2 macrophages | [52,54] |
| d | Normal fibroblasts are transformed into CAFs | [45,46] |
| e | M2 macrophages repolarize to M1 macrophages | [60,63] |
| f | M2 macrophages boost proliferation of breast cancer tumor cells | [4,56] |
| g | CAFs boost proliferation of breast cancer tumor cells | [66] |
| h | CTLs boost polarization of M2 macrophages to M1 macrophages | [62] |
| i | M2 macrophages naturally die | [115] |
| j | Breast cancer tumor cells proliferate | [116,117] |
| k | CAFs naturally die | [118] |
| l | Breast cancer tumor cells induce macrophage polarization to M2 | [4,49,58] |
| m | Breast cancer tumor cells induce transformation of normal fibroblasts to CAFs | [47,66] |
| n | M2 to M1 macrophage repolarization boosts proliferation of CTLs | [88] |
| o | M2 macrophages suppress CTL cytotoxicity | [74–77,119] |
| p | Breast cancer tumor cells boost Treg recruitment | [99–101,120] |
| q | CTLs proliferate | [92] |
| r | Breast cancer tumor cells suppress CTL cytotoxicity | [89] |
| s | Tregs are recruited | [92] |
| t | CTLs kill breast cancer tumor cells | [73] |
| u | Breast cancer tumor cells naturally die | [72] |
| v | CTLs naturally die | [92] |
| w | Tregs suppress CTL cytotoxicity | [85] |

| x | Tregs naturally die | 92 |
|---|---|---|
| y | Tregs suppress CTL proliferation | 90,91 |

**Table 2.** Brief descriptions of the mechanisms included in **Fig. 1**. CAFs = cancer-associated fibroblasts, CTLs = cytotoxic T lymphocytes ($T_C$), Tregs = regulatory T cells ($T_R$).

## Initial Value Calculations

We selected the initial cell population values to represent approximate average magnitudes of immune and stromal cell populations within a Stage II TNBC tumor rather than precise patient-specific values. Because quantitative measurements of absolute immune cell densities in human TNBC tumors are limited and heterogenous, we combined estimates from single-cell transcriptomics, histopathological studies, tumor geometry, and reported cell size ranges to generate biologically-plausible starting values. These initial values are intended to establish reasonable baseline scales for simulation rather than exact biological measurements. They are summarized in **Table 3.**

We used analysis results from the single cell transcriptomics article by Guo et. al, specifically Figure 8A, to estimate the immune cell compositions within TNBC tumor samples.[121] We estimated an average cell proportion of 10% CD8+ T cells, 4% Tregs, and 12% M2 macrophages. CAFs were not included in this study, so we use the fact that Sun et al. measured a median percentage of α-SMA-positive stroma cells to be 69% .[122] "Stroma-intermediate primary breast carcinoma" is quantified as 50% stroma.[123] Therefore, we estimated the percentage of a TNBC tumor to be $(.69)(.5) = 34.5\%$ CAFs. To calculate relative densities of immune cells within a Stage II TNBC tumor, we used the fact that Stage II tumors are diagnosed beginning at a size of 2 cm,[124] where 2 cm represents the longest diameter of the tumor. Since many breast cancers tend to be discoidal,[125] we used an ellipsoid formula to estimate the volume of a Stage II tumor to be $\frac{4}{3}\pi(1)(1)(0.5) = 2.0944 \approx 2.09\,\mathrm{cm}^3$ assuming the shorter diameter is half the length of the longest. For each cell type, we took a percentage of this volume, and then converted from volume to density using the mass of each cell type.

Limitations of these initial value calculations include the fact that the analysis performed on only ten tumor samples, that Guo et al. did not report the stage of each TNBC tumor, and that two TNBC patients had been treated before. Therefore these initial values are an estimate of average magnitude rather than an exact report of immune cell composition. Additionally, TNBC tumors are known to be heterogeneous, so initial values of immune cells, even at a collective "Stage II" timeline, certainly vary by patient. After calculating a starting estimate using the process outlined above, we tweaked our initial values using our own clinical expertise.

$M2_0$: Using the transcriptomic analysis from Guo et al., we estimated M2 macrophages to be 12% of total tumor volume,[121] implying $0.251328\,\mathrm{cm}^3$ of the initial tumor volume is M2 macrophages. Monocytes can have a diameter up to 22 microns,[126] which is equivalent to 0.0022 cm. Since monocytes are typically spherical, we estimate their volume to be $5.57528 \times 10^{-9}\,\mathrm{cm}^3$ using a spherical volume formula, so that the total number of M2 macrophages within the tumor is $\frac{0.251328\,\mathrm{cm}^3}{5.57528 \times 10^{-9}\,\mathrm{cm}^3} = 4.50789915484 \times 10^7$.
To get an average density, we assume the M2 macrophages are uniformly distributed throughout the tumor, so we calculate $\frac{45078991.5484}{2.0944\,\mathrm{cm}^3} = 2.15235826721 \times 10^7\,\mathrm{cm}^{-3}$
or approximately $21500\,\mathrm{cells/mm}$ as a starting estimate. This initial value was not subsequently adjusted.

$F_0$: Using the estimates from Sun et al. and Hagenaars et al.,[122,123] we estimated CAFs to be 34.5% of total tumor volume, implying 0.722568 $\mathrm{cm}^3$ of the initial tumor volume is CAFs. Since CAFs in our model are measured as total volume, we took 34.5% of initial tumor volume, so 0.345*2.0944 = 0.722568 $\mathrm{cm}^3$ or approximately 722 $\mathrm{mm}^3$ as a starting estimate for the initial volume of CAFs. This initial value was not subsequently adjusted.

$B_0$: Since a Stage II diagnosis means tumors with diameter no more than 2 cm, and since as mentioned previously most tumors are discoidal, we calculate an initial tumor volume of $\frac{4}{3}\pi(1)(1)(0.5) = 2.0944 \approx 2.09$, $\mathrm{cm}^3$ or 2090 $\mathrm{mm}^3$. This value was not subsequently adjusted.

$T_{C0}$: Using the transcriptomic analysis from Guo et al.,[121] we estimated CTLs to be 10% of total tumor volume, implying 0.20944 $\mathrm{cm}^3$of the initial tumor volume is comprised of CTLs. T lymphocytes have an average diameter of 9.5 microns,[127] which is equivalent to 0.00095 cm. Since T lymphocytes are spherical, we estimate their volume to be $4.4892 \times 10^{-10}\,\mathrm{cm}^3$, and the total number of CTLs within the tumor to be $0.20944\,\mathrm{cm}^3/(4.4892 \times 10^{-10}\,\mathrm{cm}^3) = 4.6654 \times 10^8$To get an average density, we assume the CTLs are uniformly distributed throughout the tumor, so we calculate $4.66541922837 \times 10^8/2.0944\,\mathrm{cm}^3 = 2.22756838635 \times 10^8\,\mathrm{cm}^{-3}$, or $2.22756838635 \times 10^5\,\mathrm{mm}^{-3}$. The 10% statistic was taken from Guo et al.'s measurement of CD8+ T cells within TNBC tumors, which may not reflect the proportion of active CTLs within TNBC tumors, which is likely less. We estimated 30% of these CD8+ T cells may be active,[128] for an initial value of $0.30 \times (2.22756838635 \times 10^5) = 66827.0515905$ or approximately 66,800.

$T_{R0}$: Using the transcriptomic analysis from Guo et al.,[121] we estimated Tregs to be 4% of total tumor volume, implying $0.083776\,\mathrm{cm}^3$ of the initial tumor volume is Tregs. We assume that Tregs, like CTLs, have an average diameter of 9.5 $\mu$,[127] which is equivalent to 0.00095 cm. Since T lymphocytes are spherical, we estimate their volume to be $4.4892 \times 10^{-10}\,\mathrm{cm}^3$ so that the total number of Tregs within the tumor is $0.083776\,\mathrm{cm}^3/(4.4892 \times 10^{-10}\,\mathrm{cm}^3) = 1.86616769135 \times 10^8$ . To get an average density, we assume the Tregs are uniformly distributed throughout the tumor, so we calculate $1.86616769135 \times 10^8/(2.0944\,\mathrm{cm}^3) = 8.9102735454 \times 10^7\,\mathrm{cm}^{-3}$. We adjusted this value to $1.09 \times 10^4\,\mathrm{cm}^{-3}$ based on our clinical authors' input.

| **Initial Values** | | | |
|---|---|---|---|
| **Cell** | **Value** | **Units** | **References** |
| $M_2$ | 21,500 | cells/mm$^3$ | 121,126 |
| $F$ | 722 | mm$^3$ | 122,123 |
| $B$ | 2,090 | mm$^3$ | 124,125 |
| $T_C$ | 66,800 | cells/mm$^3$ | 121,127 |
| $T_R$ | 10,900 | cells/mm$^3$ | 121,127 |

**Table 3.** Initial values used in simulations of the triple-negative breast cancer tumor-immune model.

We emphasize that TNBC tumors are highly heterogenous, and immune cell infiltration varies substantially between patients. Accordingly, these initial values should be interpreted as representative averages. After deriving initial estimates using the procedures described above, values were adjusted when necessary based on clinical expertise to ensure biological plausibility. The qualitative behavior of the model is not intended to depend on exact initial values but rather on the structure of interactions and relative rates along modeled mechanisms.

## Parameter Calculations

In this section, we detail the calculations used to estimate the parameters in our model. The parameter estimates and references are gathered in **Table 4**, for convenience. There are some general principles that we used for various types of parameters. To calculate the alpha parameters, we searched for experimental data that yielded the rates of growth of a population with and without the effect of interest in the context of TNBC. Then, we could take a ratio of these rates, to find a fold increase or decrease, depending on the parameter. We systematically estimated beta parameters by taking the corresponding cell population's initial value and multiplying it by 4; this is further explained in the Beta Parameters section. For proliferation rates, we found doubling times and used the formula $p = \frac{\ln 2}{d}$ where $p$ is proliferation rate and $d$ is doubling time. For some death rate constants, we found average half-life estimates ($t_{\frac{1}{2}}$) and used $\delta = \frac{\ln 2}{t_{\frac{1}{2}}}$. For other death rate constants, we found lifespan estimates and used $\delta = \frac{1}{\text{lifespan}}$. Carrying capacities were estimated as a fraction of total tumor carrying capacity, which was found through literature, and then converted to appropriate units. Recruitment rates and polarization rates were estimated from literature when possible and estimated by clinical authors' input when not. Parameters that were not calculable with literature are labeled as "estimated", to represent the estimation from clinical authors' expertise. For some parameters, we only had access to data in a setting other than TNBC. A future endeavor related to this work is gathering data to parameterize our model in the TNBC setting. Further limitations are detailed in the Discussion section.

## Alpha Parameters

Parameters of the form $\alpha_{XY}$ represent a maximum boost or inhibition that cell population $X$ can have on a mechanism from cell population $Y$. When possible, we calculated each $\alpha_{XY}$ by using experimental results that documented both the efficacy of the mechanism from $Y$ alone and the efficacy of the mechanism from $Y$ with the presence of $X$. A ratio of these efficacies represents $\alpha_{XY}$, the efficacy fold increase or decrease due to population $X$. Because each boosting and inhibition term looks like $(1 \pm \frac{\alpha_{XY} X}{\beta_{XY} + X})$, $(1 \pm \alpha_{XY})$ is the limit of this term as $X$ gets larger and larger, so in fact $(1 \pm \alpha_{XY})$ is the maximum boost or inhibition, and thus in a boosting case, $\alpha_{XY}$ is obtained by taking the ratio of efficacies and subtracting 1. In an inhibition case, $\alpha_{XY}$ is obtained by subtracting the ratio of efficacies from 1. These conventions are used throughout the parameter calculations section.

$\alpha_{F2}$: Cohen et al. studied how Chi3L1 recruits macrophages and how it induces an M2-like phenotype in breast cancer.[43] Figure 4L in Cohen's study shows the fold change of the ratio of CD206/CD86 expression between macrophages mixed with recombinant Chi3L1 and serum-free macrophages (SFM). We estimated the fold change to be 2.2 from this diagram, resulting in $\alpha_{F2} = 2.2 - 1 = 1.2$.

$\alpha_{C2}$: Sharma et al. studied mechanistic interactions between intratumoral effector T cells and tumor-associated macrophages.[62] The authors show CTL activity can be restored with a combination of anti-CTLA-4 and a vaccine that activates the ICOS pathway (IVAX). Since $\alpha_{C2}$ represents the maximum effect CTLs can have on repolarizing M2 macrophages to M1 macrophages, we use their control setting (black control bar in Figure S3D of Sharma et al.) to estimate the efficacy of M2 repolarizing to M1 in absence of CTLs, and their setting with anti-CTLA-4 and IVAX treatment (blue bar in Figure S3D of Sharma et al.) to estimate the efficacy of repolarization in the presence of CTLs. We estimate the black bar (control) in Figure S3D in Sharma et al. to be 48% and the blue bar (anti-CTLA-4 + IVAX) to be 12%,[62] giving us a fold change of 0.48/0.12 = 4. Thus, $\alpha_{C2} = 4 - 1 = 3$.

$\alpha_{2B}$: Tu et al. studied how M2 macrophages contribute to breast cancer cell proliferation.[129] We used Figure 1B in Tu's study to calculate the maximum boosting effect M2 macrophages can have on triple-negative breast cancer cell proliferation. Because we are estimating the maximum effect, we chose to use the timepoint of 48 hours, when the difference between the proliferation rate in the control (black line) and the proliferation rate in the setting with the supernatant medium of M2 macrophages (blue line) is the largest. We took 2.9/2.2 as the ratio of proliferation of breast cancer cells with M2 macrophages to proliferation of breast cancer cells with no macrophages to get 1.318. Thus, $\alpha_{2B}$ = 1.318 - 1 = 0.318.

$\alpha_{FB}$: Takai et al. studied CAFs in human TNBC and how they promote tumor growth.[46] We used Figure 2B to approximate the fold change of tumor growth in the presence CAFs versus tumor growth in the absence CAFs, assuming the loss of TNBC is negligible in each setting. We take the widest gap, at Day 25, and estimate the tumor volume with CAFs (gray curve) to be 2100 $\mathrm{mm}^3$, and estimate the tumor volume without CAFs (black curve) to be 800 $\mathrm{mm}^3$, which yields a fold change of 2100/800 = 2.625. Thus, $\alpha_{FB} = 2.625 - 1 = 1.625 \approx 1.63$.

$\alpha_{CB}$: Khazen et al. studied how CTL population density can affect tumor killing efficacy in melanoma tumors.[130] We used Figure 5 in their study to estimate the maximum cytotoxic effect CTLs can have on breast cancer tumor cells. We used the widest gap, at Day 7, to estimate a fold change of tumor size with no injections of CTLs versus tumor size with 3 injections of CTLs. At Day 7, we estimate the tumor size with no injections (black curve) to be 150 $\mathrm{mm}^3$, and the tumor size with 3 injections (green curve) to be 5 $\mathrm{mm}^3$. Then to obtain a fold change, we take 150/5 = 30, so $\alpha_{CB} = 30 - 1 = 29$.

$\alpha_{2CB}$: Le Mercier et al. studied the V-domain immunoglobulin suppressor of T-cell activation (VISTA) checkpoint in mice with melanoma.[107] Figure 1A in their article shows tumor growth in a control population and tumor growth in a population treated with 13F3, a VISTA inhibitor. Because the VISTA inhibitor will inhibit polarization of M1 macrophages to M2 type,[131] we consider the black curve (control) to represent tumor growth where M2 macrophages are present, inhibiting CTL activity maximally. We consider the red curve (13F3) to represent tumor growth without any M2 inhibition of CTLs killing tumor cells. Thus, to calculate the maximum potential suppressive activity of M2 on CTLs, we chose to take the widest gap between the black and red curves, at Day 18. We estimated the untreated tumor size at Day 18 to be 91 $\mathrm{mm}^2$ and the 13F3-treated tumor size at Day 18 to be 22 $\mathrm{mm}^2$. Therefore, we calculated the ratio of the CTL killing of tumor cells with no M2 inhibition of CTLs to the maximal M2 inhibition of CTL killing of tumor cells to be 22/91 = 0.24175824175. The number 0.24175824175 corresponds to the maximum inhibitory effect M2 macrophages can have on CTLs, so to reflect this in the model we take $\alpha_{2CB} = 1 - 0.24175824175 = 0.75824175825 \approx 0.758$.

$\alpha_{BCB}$: At the end of Equation (3), the factors

$$\overbrace{(1 + \frac{\alpha_{CB} T_C}{\beta_{CB} + T_C}}^{\text{t}} \overbrace{[(1 - \frac{\alpha_{2CB} M_2}{\beta_{2CB} + M_2})}^{\text{o}} \overbrace{(1 - \frac{\alpha_{BCB} B}{\beta_{BCB} + B})}^{\text{r}} \cdots$$

represent the combined effects of **arrows t**, **o**, and **r** on the death of breast cancer cells. Shi et al. found the average percent of CD8+ T cells producing granzyme in healthy donors, which we use as a proxy to measure cytotoxic activity, to be approximately 30% in Figure 1G in their paper,[128] and Abudula et al. showed in Figure 5F in their paper that the percent of CD8+ T cells producing granzyme under the suppressive effects of tumor cells and macrophages is approximately 1.3%.[119] This implies the total reduction effect of both tumor cells and macrophages combined can be estimated by a fold change of $0.013/0.3 \approx 0.04333333333$. If we assume this is the largest possible suppression, then we have

$$\overbrace{(1-\alpha_{2CB})}^{o}\overbrace{(1-\alpha_{BCB})}^{r} = 0.04333333333$$

and since we have

$$\overbrace{(1-\alpha_{2CB})}^{o} = 0.24175824175$$

(see $\alpha_{2CB}$), then

$$\overbrace{(1-\alpha_{BCB})}^{r} = 0.04333333333/0.24175824175 = 0.17924242423,$$

and

$$\alpha_{BCB} = 1 - 0.17924242423 = 0.82075757577 \approx 0.821.$$

$\alpha_{RCB}$: Chen et al. studied how Tregs can suppress CTL cytotoxicity through TGF-β in a mouse model of colon carcinoma.[132] We used Figure 4b to estimate about 55% of naive CD8 cells are capable of lysis versus 25% of CD8 cells are capable of lysis when they are co-cultured with Tregs. Therefore we calculate the fold change of CTL activity without Tregs versus CTL activity with Tregs to be 0.25/0.55 = 0.455, so $\alpha_{RCB} = 1 - 0.455 = 0.545$.

$\alpha_{2C}$: The $\alpha_{2C}$ parameter is used to represent the maximum boosting effect M1 macrophages, modeled by the loss of M2 macrophages, can have on CTL proliferation. Kaneda et al. show that a PI3Kγ inhibitor can repolarize TAMs from M2 to M1 in breast cancer.[133] Using Figure 3c, we estimated a fold change of CD8+ T cells' ability to proliferate in the presence of macrophages with PI3K (green triangles) versus without PI3K (dark blue circles) to be 80/10 = 8. Therefore, $\alpha_{2C} = 8 - 1 = 7$.$\alpha_{2C} = 8 - 1 = 7.$

$\alpha_{RC}$: Feinerman et al. studied regulatory T cells, effector T cells, IL-2, and interactions among these.[134] We use the IL-2 concentration measured in Figure 6b as a proxy for CTL proliferation potential. We estimated that at 61 hours the concentration of IL-2 in a setting with weak effector T cells in the presence of Tregs was 1.35 pM, compared to the concentration of IL-2 in a setting with weak effector T cells in the absence of Tregs, which we estimated to be 25 pM. Thus, the fold decrease in IL-2 due to the presence of Tregs is $\frac{1.35}{25} = 0.054$, so the maximum inhibition Tregs can have on CTL proliferation is $1 - 0.054 = 0.946.$

$\alpha_{2R}$: Semba et al. studied JNK-active TNBC clusters in the TNBC TME in mice.[135] They found JNK promotes an immunosuppressive TME via macrophage-derived CCL2. 80% of the CCL2 expressing cells are TAMs. Therefore, in Figure 1c, we assume the difference between the estimated fraction of Tregs in pJNK-low tumors and pJNK-high tumors is due to M2 macrophages. The estimated fraction of Tregs in pJNK-high tumors is 2.5, and the estimated fraction of Tregs in pJNK-low tumors is 1.7. Therefore, we estimate the fold change of Treg recruitment in the presence of M2 macrophages versus in the absence M2 macrophages to be 2.5/1.7 = 1.47058823529, so
$\alpha_{2R} = 1.47058823529 - 1 = 0.47058823529 \approx 0.471.$

$\alpha_{BR}$: Zahran et al. studied the accumulation of Tregs in patients with TNBC.[94] We calculated the maximum boosting effect TNBC can have on recruiting Tregs using Table 2 in their paper. The authors found 6.6% of malignant tissue was CD4+CD25+highFoxP3+Tregs versus 3.2% in normal tissue. We can use these percentages to calculate a $0.066/0.032 = 2.0625$ fold change between Treg recruitment in the presence of TNBC versus Treg recruitment in normal tissue. Therefore, $\alpha_{BR} = 2.0625 - 1 = 1.0625 \approx 1.063$.

## Beta Parameters

Parameters of the form $\beta_{XY}$ represent the threshold (cell X population size) for half of the maximum effect by cell $X$ on a cell $Y$ mechanism. To obtain estimates of the $\beta_{XY}$ parameters, we multiplied each initial value by 4, assuming that each cell population would need to double twice to achieve this threshold. This assumption was made to achieve simulations that matched pre-specified expected behavior (i.e., unit tests for model behavior).

$\beta_{2C}$: We calculated $\beta_{2C}$ by taking $4 \times M_{2_0}$, and then dividing by $\gamma_2 = 0.005$ to account for unit conversion, to get our final estimate of $17{,}200{,}000 \text{ cells/mm}^3$.

$\beta_{2X}$: For all other betas associated with M2 macrophages, we calculated $4 \times M_{2_0} = 4 \times 21500 = 86000 \text{ cells/mm}^3$.

$\beta_{FX}$: For any beta associated with CAFs, we calculated $4 \times F_0 = 4 * 722 = 2888 \approx 2890$.

$\beta_{BX}$: For any beta associated with TNBC cells, we calculated $4 \times B_0 = 4 \times 2090 = 8360 \text{ mm}^3$.

$\beta_{CX}$: For any beta associated with CTLs, we calculated $4 \times T_{C0} = 4 * 66800 = 267200$.

$\beta_{RX}$: For any beta associated with Tregs, we calculated $4 \times T_{R0} = 4 * 10900 = 43600$.

## Carrying Capacities

$K_F$: The carrying capacity for CAFs is not measured directly in the literature. However, as detailed in the initial value calculations, we estimated that up to 34.5% of an average TNBC tumor is comprised of CAFs. To ensure the carrying capacity of CAFs that we selected would be applicable even in larger tumors, $K_F$ was taken to be 50% of $K_B$. Hence, $K_F = 0.5 \times 262 = 131 \text{ cm}^3 = 131000 \text{ mm}^3$.

$K_B$: This parameter was calculated using data from a study by Chen et al. that found a TNBC tumor with a 10 cm diameter,[136] which is the largest that has been reported in the literature. Since breast cancer tumors tend to be discoidal,[125] we estimated the maximum possible volume of a TNBC tumor to be

$$K_B = \frac{4}{3}\pi(5)(5)(2.5) = 261.79939 \approx 262 \text{ cm}^3 = 262000 \text{ mm}^3,$$

where 5 is the radius of the longer directions and 2.5 is the radius of the shorter one.

$K_C$: Using the fact that $K_B = 262000 \text{mm}^3$, and that CTLs make up roughly 10% of a TNBC tumor, about 26,200 $\text{mm}^3$ can contain $T_C$. As calculated in the initial value section, $T_C$ has an estimated volume to be $4.489 \times 10^{-10} \text{ cm}^3$ or $4.489 \times 10^{-7} \text{ mm}^3$. This means $\frac{2{,}620{,}013.09}{4.489 \times 10^{-7}} = 58{,}364{,}891{,}958.1$ total $T_C$ can fit in that size of a tumor. Dividing by 262,000 $\text{mm}^3$ again yields 222,766.763199 $\text{cells/mm}^3$ or approximately 223,000 $\text{cells/mm}^3$ for $K_C$.

## Source and Proliferation Parameters

$\alpha_{B2}$: Leonard et al. measured macrophage migration toward mouse model breast cancer cells in a migration chamber.[137] We used Figure 2b in their article to estimate that 200 macrophages, as noted by the conditioned 4T1 medium column of the control, migrated toward tumor cells over a 48 hour period. Assuming a linear migration rate, this amounts to 100 macrophages per day. This study used 8000-10000 breast cancer tumor cells, so we estimate the recruitment rate in our model to be higher, due to the number of breast cancer cells in our setting being orders of magnitude higher. We therefore estimated $\alpha_{B2}$ to be 10,000 macrophages/day.

$\sigma_F$: Longitudinal data on transformation rates of normal fibroblasts into CAFs is limited. We assume that CAF transformation is much slower than tumor cell proliferation.[138] To reflect this assumption, we estimated $\sigma_F = 0.1 p_B = 0.000998$.

$p_B$: was calculated using a study by Lee et al. that found an average tumor growth rate of 1.003%/day in a group of 67 patients with triple-negative breast cancer.[116] To convert this to a 1/day rate constant, we use an exponential growth formula

$$1.01003 = e^{p_B}$$

so we have $p_B = 0.00998$.

$p_C$: de Boer et al. (Table III) found proliferation rates of CD8+ T cells for six different viral epitopes by fitting a model to data from mice infected with lymphocytic choriomeningitis virus. CD8+ T cell clonal expansion is rapid and vast during a viral infection. However, clonal expansion rate in the setting of the immunosuppressive breast cancer TME is likely to be smaller. Hence, we conservatively selected the smallest reported proliferation rate constant of 1.13/day for the nominal value in our setting.[139]

$\sigma_R$: Shanahan et al. studied Treg recruitment in a mouse model of lung cancer.[140] Using Figure 5A, we estimated 6700 Tregs were recruited to the lung of a tumor-bearing mouse in 16 hours, or 2/3 days. We estimated that a mouse's lung has a total volume of 1 milliliter,[141] and we used Figure 5G to estimate 1.85% of the total lung is tumor. This implies that $1000\mu\mathrm{L} \times 0.0185 = 18.5\mu\mathrm{L}$ is the total volume of the tumor used in the control of Figure 5. We thus calculated a source rate as follows:

$$\frac{6700}{(18.5)(\frac{2}{3})} = 543.2433 \approx 543$$

.

## Loss Parameters

$\delta_2$: This parameter was estimated using Eftimie and Barelle's model of immune responses in lung cancer that was parameterized with mouse data.[142] The natural death rate of macrophages fell in a range between 0.09 and 0.14, and we chose 0.1155, the midpoint of this range.

$\delta_F$: Weissman-Shomer and Fry found an average lifespan of chick embryo fibroblasts to be 57 days.[143] Since TNBC produces chronic inflammation, we expect CAFs to have a longer lifespan, so we took 0.1 times 1/57, which gives 0.00175.

$\delta_B$: Mahacek et al. showed that SUM-44PE cells, which are cells from a human ER+ breast cancer cell line, had a proliferative lifespan of at least two years, or 730 days, in vitro.[72] The exact length of proliferative lifespan was not measured, so we used 730 days as a lower bound for the lifespan of TNBC. Note that the death rate is calculated by taking the reciprocal of the lifespan. Taking the reciprocal of a lower bound will yield an upper bound. A lifespan of 730 days would give a death rate constant of 1/730 = 0.00134, an upper bound for the death rate constant for TNBC; we estimated that $\delta_B$ = 0.0005.

$\delta_C$: de Boer et al. found a half-life of 41 hours for activated CD8+ T cells.[139] We converted this to an average death rate by dividing $\ln(2)$ by half-life; we calculated

$$\delta_C = \frac{\ln 2}{\frac{41}{24}} = 0.406$$
.

$\delta_R$: Vukmanovic-Stejic et al. used deuterium labeling to monitor disappearance of Tregs in healthy volunteers.[95] The authors found that Tregs have a half-life of about 11 days, which yields an average death rate constant of $\delta_R = 0.0630$.

Repolarization and Proportionality Parameters

$\gamma_2$: Xiao et al. induced M2 to M1 repolarization using targeted codelivery of STAT6 inhibitor and IKKβ siRNA to mice in a subcutaneous breast cancer model.[106] Mice in the control group (treated only with phosphate-buffered saline) had 1.81% of total tissue as M2-like TAMs on Day 8, whereas mice treated with combination therapy group (ST-AS&Si) had 0.33%. We assumed that ST-AS&Si treatment causes more $M_2$ to $M_1$ polarization than usual. Thus we used $\frac{0.0033}{0.0181} \approx 0.1823$ as the lower bound for the fold change of number of M2 macrophages between a treated and untreated setting. In our differential equation for the rate of change of $M_2$, the expression $-\gamma_2 M_2$ represents the rate of $M_2$ to $M_1$ polarization. Thus we used the following exponential decay formula to calculate an upper bound for $-\gamma_2 M_2$ represents the rate of $M_2$ to $M_1$ polarization. Thus we used the following exponential decay formula to calculate an upper bound for $\gamma_2$:

$$e^{-8\gamma_2} = 0.1823$$

where 8 is chosen because macrophages were measured on Day 8. Solving for $\gamma_2$ gives 0.2128 as an upper bound for the basal M2 to M1 repolarization rate. Notably, this estimate reflects a treated state. In an untreated state, the basal rate is likely to be much lower. We estimated the basal rate in an untreated setting to be 0.005.

| Model Parameters | | | | | | | |
|---|---|---|---|---|---|---|---|
| # | Symbol | Description | Label | Value | Units | Sources | Setting |
| 1 | $\alpha_{B2}$ | Maximum source rate of tumor-associated M2 macrophages | **c,l** | 10,000 | cells/mm$^3$*day | [137] | Mouse model of TNBC |
| 2 | $\beta_{B2}$ | Threshold for half of maximum effect of TNBC cells on M2 source<br><br>Threshold for half of maximum effect of TNBC cells inducing M2 | **c,l** | 8,360 | mm$^3$ | estimated | |
| 3 | $\alpha_{F2}$ | Maximum boost by CAFs on M1 to M2 polarization | **a** | 1.2 | unitless | [43] | Mouse model of breast cancer |
| 4 | $\beta_{F2}$ | Threshold for half of maximum boost by CAFs on M1 to M2 polarization | **a** | 2,890 | mm$^3$ | estimated | |
| 5 | $\gamma_2$ | M2 to M1 polarization rate constant | **e** | 0.005 | 1/day | [106] | Mouse model of TNBC |
| 6 | $\alpha_{C2}$ | Maximum boost by CTLs on M2 to M1 polarization | **h** | 3 | unitless | [62] | Mouse model of melanoma |
| 7 | $\beta_{C2}$ | Threshold for half of maximum boost by CTLs on M2 to M1 polarization | **h** | 267,200 | cells/mm$^3$ | estimated | |
| 8 | $\delta_2$ | M2 loss rate constant | **i** | 0.116 | 1/day | [142] | Non-small cell lung cancer math model parameterize d with mouse data |

| 9 | $\sigma_F$ | CAF source rate constant | **d,m** | 0.000998 | 1/day | estimated | |
|---|---|---|---|---|---|---|---|
| 10 | $K_F$ | CAF carrying capacity | **d,m** | 65,500 | $mm^3$ | [121–123] | Human patients with breast cancer (multiple subtypes) |
| 11 | $\delta_F$ | CAF loss rate constant | **k** | 0.00175 | 1/day | [143] | Normal fibroblasts in chick embryos |
| 12 | $p_B$ | TNBC cell proliferation rate constant | **j** | 0.00998 | 1/day | [116] | Human patients with TNBC |
| 13 | $K_B$ | TNBC cell carrying capacity | **j** | 262,000 | $mm^3$ | [125,136] | Human patients with TNBC |
| 14 | $\alpha_{2B}$ | Maximum boost by M2 macrophages on TNBC cell proliferation | **f** | 0.318 | unitless | [129] | Human patients with TNBC |
| 15 | $\beta_{2B}$ | Threshold for half of the maximum boost by M2 macrophages on TNBC cell proliferation | **f** | 86,000 | $cells/mm^3$ | estimated | |
| 16 | $\alpha_{FB}$ | Maximum boost by CAFs on TNBC cell proliferation | **g** | 1.63 | unitless | [46] | Mouse xenograft model of TNBC |
| 17 | $\beta_{FB}$ | Threshold for half of the maximum boost by CAFs on TNBC cell proliferation | **g** | 2,890 | $mm^3$ | estimated | |
| 18 | $\delta_B$ | TNBC cell loss rate constant | **u** | 0.0005 | 1/day | [72] | Human patients with TNBC |
| 19 | $\alpha_{CB}$ | Maximum cytotoxic effect by CTLs on | **t** | 29 | unitless | [130] | Mouse model of |

| | | TNBC cells | | | | | melanoma |
|---|---|---|---|---|---|---|---|
| 20 | $\beta_{CB}$ | Threshold for half of the maximum cytotoxic effect by CTLs on TNBC cells | **t** | 267,200 | cells/mm$^3$ | estimated | |
| 21 | $\alpha_{2CB}$ | Maximum suppression by M2 macrophages on CTL cytotoxic activity | **o** | 0.758 | unitless | [107] | Mouse model of melanoma |
| 22 | $\beta_{2CB}$ | Threshold for half of the maximum suppression by M2 macrophages on CTL cytotoxic activity | **o** | 86,000 | cells/mm$^3$ | estimated | |
| 23 | $\alpha_{BCB}$ | Maximum suppression by TNBC cells on CTL cytotoxic activity | **r** | 0.821 | unitless | [119,128] | Healthy human subjects; mouse model of TNBC |
| 24 | $\beta_{BCB}$ | Threshold for half of the maximum suppression by TNBC cells on CTL cytotoxic activity | **r** | 8,360 | mm$^3$ | estimated | |
| 25 | $\alpha_{RCB}$ | Maximum suppression by Tregs on CTL cytotoxic activity | **w** | 0.545 | unitless | [132] | Mouse model of colon carcinoma |
| 26 | $\beta_{RCB}$ | Threshold for half of the maximum suppression by Tregs on CTL cytotoxic activity | **w** | 43,600 | cells/mm$^3$ | estimated | |
| 27 | $p_C$ | CTL proliferation rate constant | **q** | 1.13 | 1/day | [139] | Mouse model of acute lymphocytic choriomenin gitis |
| 28 | $K_C$ | CTL carrying capacity | **q** | 223,000 | cells/mm$^3$ | estimated | |
| 29 | $\alpha_{2C}$ | Maximum boost by M1 macrophages on CTL proliferation | **n** | 7 | unitless | [133] | Mouse models of various cancers |
| 30 | $\beta_{2C}$ | Threshold for half of the maximum boost by M1 macrophages on CTL proliferation | **n** | 17,200,000 | cells/mm$^3$ | estimated | |
| 31 | $\alpha_{RC}$ | Maximum inhibition by Tregs on CTL proliferation | **y** | 0.946 | unitless | [134] | In vitro studies of cells from healthy mice |
| 32 | $\beta_{RC}$ | Threshold for maximum inhibition by Tregs on CTL proliferation | **y** | 43,600 | cells/mm$^3$ | estimated | |
| 33 | $\delta_C$ | CTL loss rate constant | **v** | 0.406 | 1/day | [139] | Mouse model of acute lymphocytic choriomenin gitis |
| 34 | $\sigma_R$ | Treg recruitment rate | **s** | 583.33 | cells/mm$^3$*day | [140] | Mouse model of lung cancer |
| 35 | $\alpha_{2R}$ | Maximum boost by M2 macrophages on Treg recruitment | **b** | 0.471 | unitless | [135] | Mouse model of TNBC |
| 36 | $\beta_{2R}$ | Threshold for the maximum boost by M2 macrophages on Treg recruitment | **b** | 86,000 | cells/mm$^3$ | estimated | |
| 37 | $\alpha_{BR}$ | Maximum boost by TNBC cells on Treg recruitment | **p** | 1.063 | unitless | [94] | Human patients with TNBC |
| 38 | $\beta_{BR}$ | Threshold for half | **p** | 8,360 | mm$^3$ | estimated | |

| | | of the maximum boost by TNBC cells on Treg recruitment | | | | | |
|---|---|---|---|---|---|---|---|
| 39 | $\delta_R$ | Treg loss rate constant | **x** | 0.0630 | 1/day | 95 | Healthy human subjects |

**Table 4. Parameter descriptions and their nominal values.** This table lists the parameter descriptions, nominal values, units, data sources, and experimental or clinical settings used to parameterize the model.

# Simulations and Sensitivity Analysis

Our simulations provide predictions for the dynamics of the TME of a TNBC tumor. All cell populations increase over time, except CTLs initially increase and then approach a lower steady-state value (**Fig. 2**). Within a few hundred days, immune cells within an untreated TNBC tumor will reach their steady-states, with CAFs still growing in number and likely contributing to metastasis and epithelial-to-mesenchymal transition (**Fig. 3**).[144] By Day 500 in our model, an untreated Stage II TNBC tumor is predicted to be close to its carrying capacity (**Fig. 3**, top right).

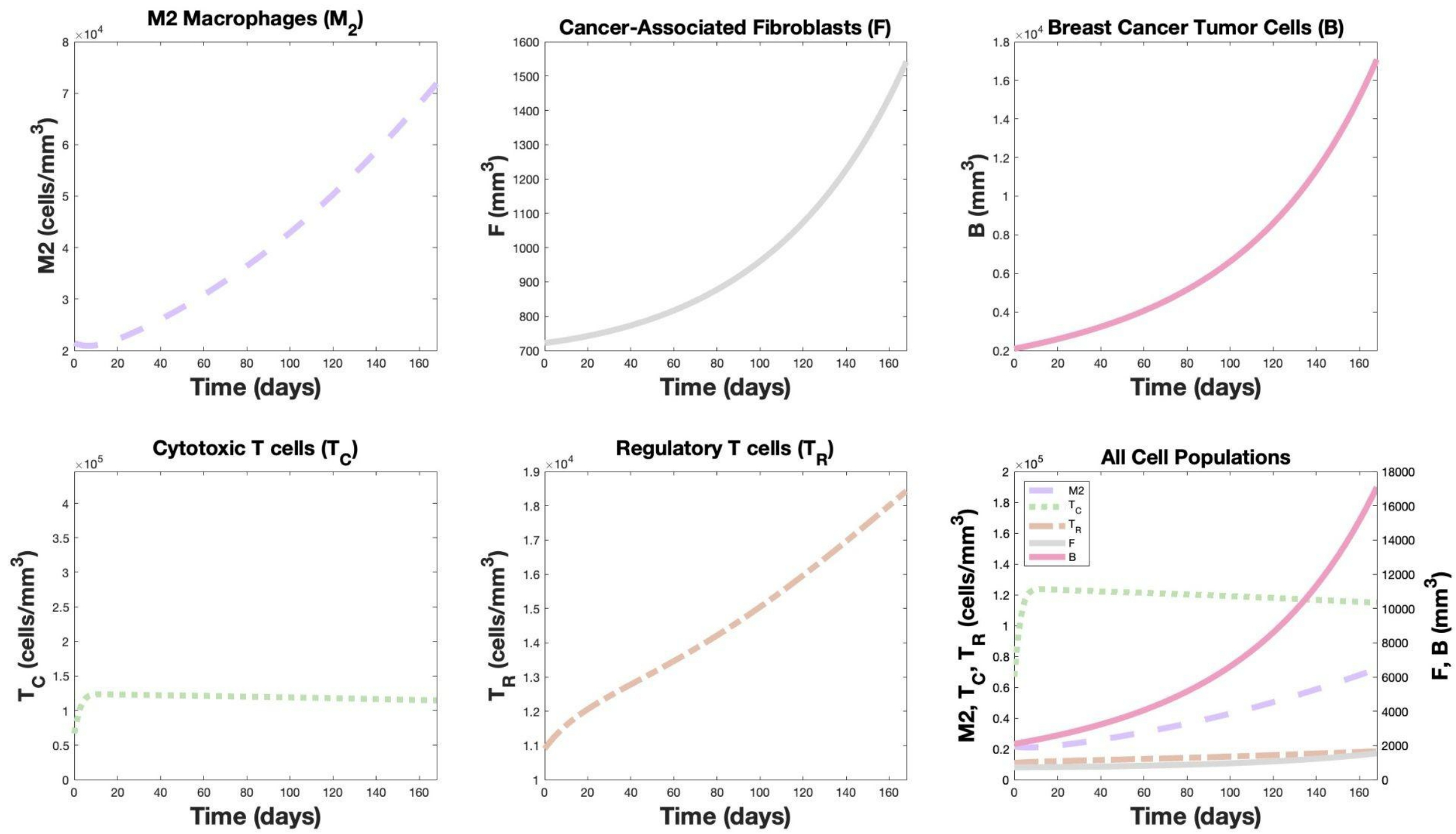

**Fig. 2. Simulated cell population levels over 168 days.**
We simulated over 168 days to match typical treatment periods. Each population is plotted in a color matching its symbol in our model diagram (M2 in dashed lavender, CAFs in solid gray, TNBC in solid pink, CTLs in dotted green, Tregs in dashed-dotted orange). The bottom right plot shows all cell populations together, where the left axis is in units of cells/mm$^3$ (for M2, CTLs, and Tregs) and the right axis is in units of mm$^3$ (for CAFs and TNBC cells). Plots were created with the ode45 function in MATLAB R2024b using nominal parameter values and initial values.

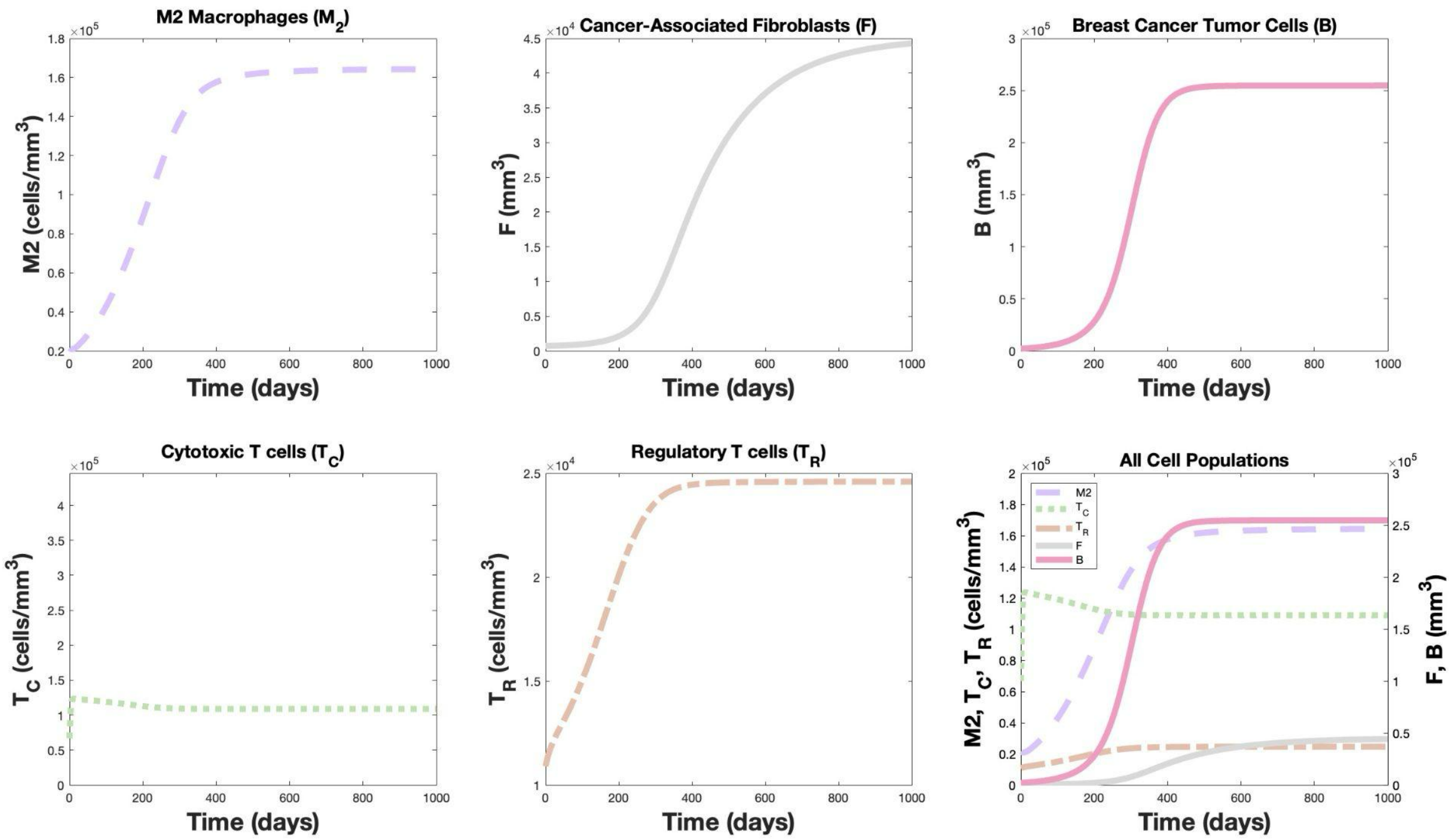

**Fig. 3. Simulated cell population levels over 1000 days.** The same plot as **Fig. 2**, but with a timeframe of 1000 days to show longer-term behavior of each cell population in our model. Each population is plotted in a color matching its symbol on our model diagram (M2 in dashed lavender, CAFs in solid gray, TNBC in solid pink, CTLs in dotted green, Tregs in dashed-dotted orange). The bottom right plot shows all cell populations together, where the left axis is in units of cells/mm$^3$ (for M2, CTLs, and Tregs) and the right axis is in units of mm$^3$ (for CAFs and TNBC cells). Plots were created with the ode45 function in MATLAB R2024b using nominal parameter values and initial values.

We performed a variance-based global sensitivity analysis using the Sobol' method.[145,146] Our model has 39 parameters. Each parameter was allotted a uniform distribution, with its range bounded below by half its nominal value, and bounded above by 1.5 times its nominal value when possible. There were three exceptions, the inhibitory alpha parameters $\alpha_{2CB}$, $\alpha_{BCB}$, and $\alpha_{RC}$, which each have a nominal value larger than 2/3. Multiplying the nominal value of $\alpha_{2CB}$ by 3/2 would result in an upper bound higher than 1, which would cause $1-\alpha_{2CB}$ to become negative for some values of $\alpha_{2CB}$, resulting in an unintended positive sign the loss rate of TNBC in Equation 3. A similar unintended sign change would occur if we multiplied the nominal values of $\alpha_{BCB}$ and $\alpha_{RC}$ by 3/2. Thus, we restricted those three parameters to have a range with an upper bound set equal to 1. We used 175,000 base samples for the sensitivity analysis, which resulted in (39+2)*(175,000) = 7,175,000 model evaluations in order to compute the Sobol' index values. Performing this type of sensitivity analysis determines the relative influence that the model parameters have on a specified quantity-of-interest (QOI). Our QOI is the volume of the TNBC tumor on Day 168, the end of a typical TNBC chemoimmunotherapy treatment period of 24 weeks.[103] We chose this QOI because we plan to use our model to compare clinical treatment regimens in silico in the future, and we wanted to determine which parameters influence the tumor size in a clinically-relevant timeframe and without any treatment, to establish a baseline. The results of our sensitivity analysis are shown in **Fig. 4**. In descending order, the top 5 most-influential parameters in our model are the proliferation rate constant for TNBC cells ($p_B$), the maximum boost by CAFs on TNBC proliferation ($\alpha_{FB}$), the threshold for half of the maximum boost by CAFs on TNBC proliferation ($\beta_{FB}$), the carrying capacity of the TNBC tumor ($K_B$), and the maximum boost by M2 macrophages on TNBC proliferation ($\alpha_{2B}$). These parameters are represented graphically in **Fig. 5**, with color highlights on the arrows representing the pathways they are part of.

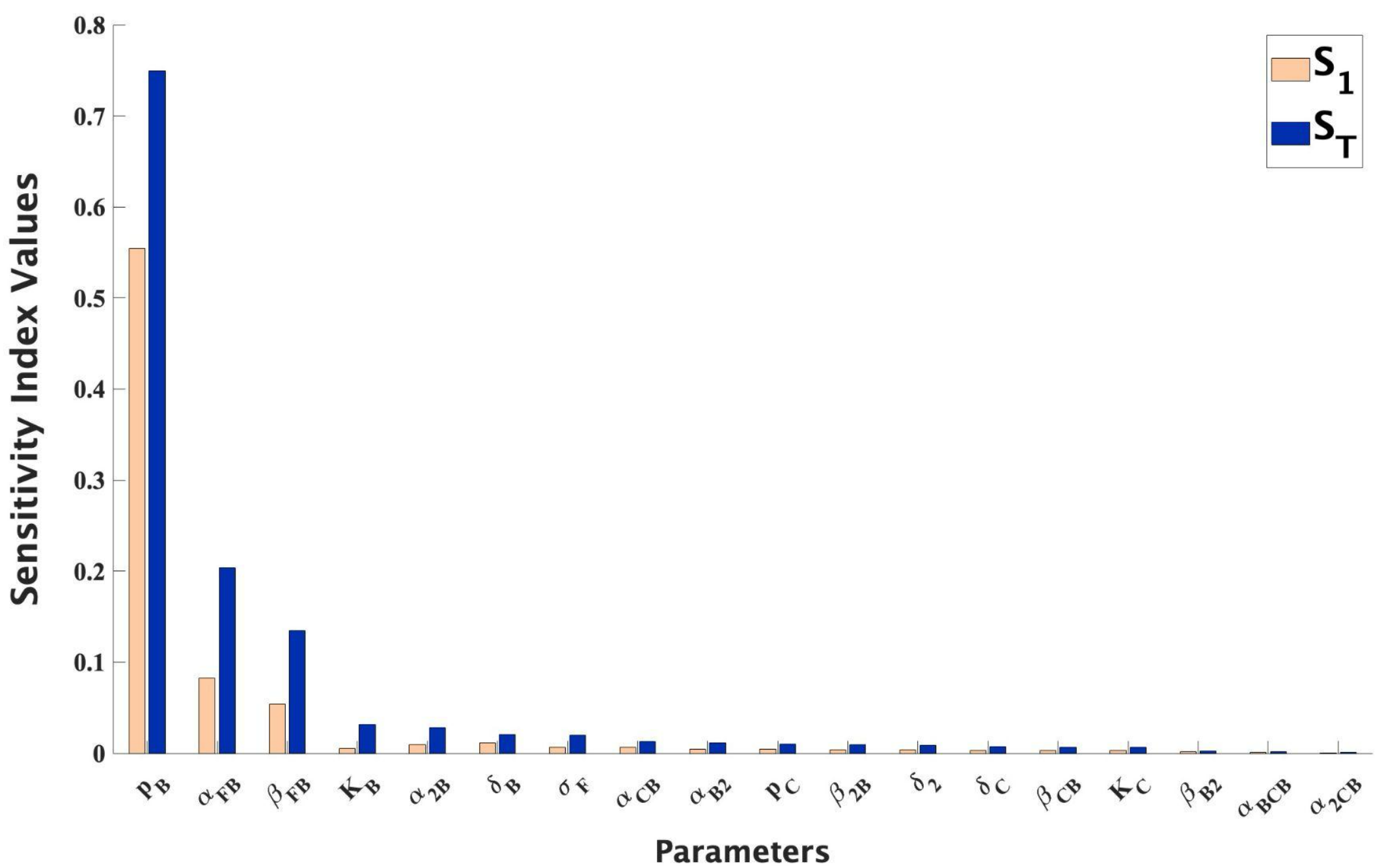

**Fig. 4. Global sensitivity analysis indices for TNBC tumor size at Day 168.** This bar chart shows the 18 parameters that are most influential on the size of the TNBC tumor on Day 168, in descending order of total-order Sobol' sensitivity index, $S_T$, shown in dark blue. Parameters not shown have a total-order Sobol' sensitivity index less than $10^{-3}$. The height of each dark blue bar is the value of the total-order Sobol' sensitivity index, $S_T$, and the height of each light orange bar is the value of the first-order Sobol' sensitivity index, $S_1$. We used code written in MATLAB R2024b to perform this Sobol' global sensitivity analysis with 175,000 base samples.

The dominance of tumor-intrinsic parameters, such as the TNBC proliferation rate constant and tumor carrying capacity, is expected given their direct role in managing tumor growth. Importantly, beyond these intrinsic factors, the sensitivity analysis reveals that stromal- and immune-mediated mechanisms, particularly boosting of tumor growth by CAFs and M2 macrophages, are among the most-influential contributors to tumor size. These findings suggest that modulation of the TME may substantially impact disease progression, and suggest biological interactions that may offer the greatest leverage for combination treatment strategies, rather than ranking parameters in isolation.

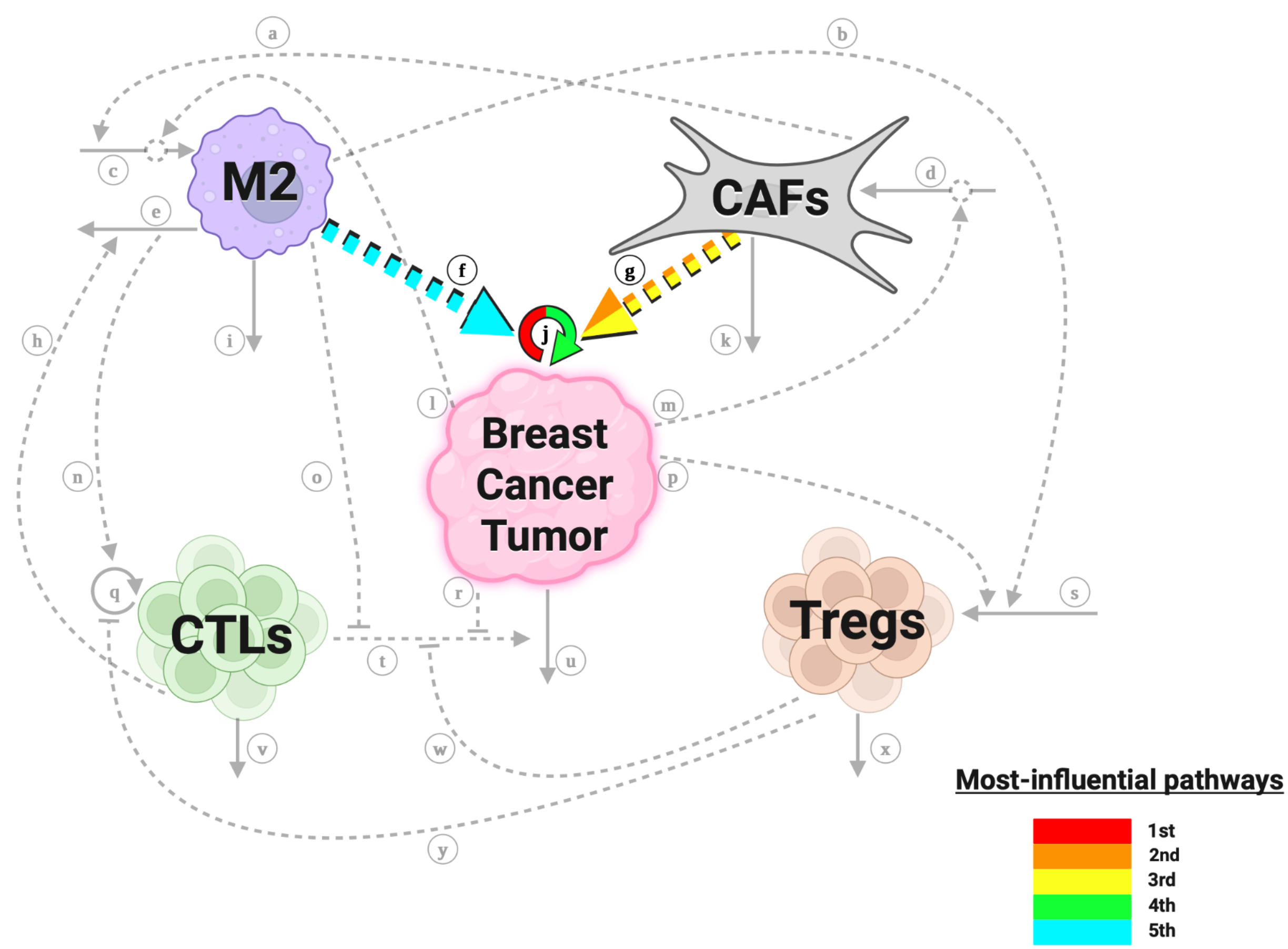

**Fig. 5. Heat map of sensitivity analysis results.** Influential arrows are highlighted in our model diagram according to a rainbow-colored heat map of descending level of influence. Red indicates the arrow that contains the most-influential parameter in our model, orange indicates the pathway with the second-most-influential parameter, etc. **Arrow j** is split into two colors, as the most-influential parameter, $p_B$, and the fourth-most-influential parameter, $K_B$, are both involved in mechanisms represented by **arrow j**. Similarly, **arrow g** is split into two colors, as the second-most-influential parameter, $\alpha_{FB}$, and the third-most-influential parameter, $\beta_{FB}$, are both involved in **arrow g.** The fifth-most-influential parameter, $\alpha_{2B}$, corresponds to **arrow f.** Figure created with Biorender.com.

We selected the 5 parameters with the largest total Sobol' indices in **Fig. 4** as the most-influential parameters. We focused on only a small subset of the parameters to ensure we are targeting the most-important mechanisms with respect to tumor burden. Also, since available data are sparse in our setting, it is computationally advantageous to plan to fit our model with respect to a few parameters, rather than dozens. The ordering of these top 5 parameters remained constant for total Sobol' indices calculated with 100,000 base samples and higher (**Table S1** in the Supplementary Material).
To qualify how well the 5 most-influential parameters capture the behavior of the full model, we generated histograms of QOI values (**Fig. 6**), with samples generated while varying all 39 parameters (shown in black) versus varying only the 5 most-influential parameters are varied (shown in orange) versus varying just the remaining 34 least-influential parameters (shown in blue). To obtain these curves, we randomly generated parameter sets using the same ranges as were used in the global sensitivity analysis. To produce the black curve, we generated 250,000 parameter sets with all 39 parameters selected from their specified distributions. For the orange curve, another 250,000 parameter sets were generated, with only the top 5 most-influential parameters allowed to vary; the remaining 34 least-influential parameters were fixed at their nominal value. In contrast, the blue curve was produced by generating another 250,000 parameter sets, where the 34 least-influential parameters were allowed to vary, and the 5 most-influential parameters were frozen at their nominal values. The similarity of the black and orange histograms indicates that if we only vary the 5 most-influential parameters, we obtain almost all of the variability present when all 39 parameters are varied. In contrast, it is apparent that varying the 34 remaining parameters captures only a very small part of the QOI-value range, and much less than varying the 5 most-influential parameters does. **Fig. S1** in the Supplementary Material shows the full histogram, as **Fig. 6** shows a truncated version for close-up viewing.

We used the two-sample Kolmogorov-Smirnov (KS) test to quantitatively compare the sets of QOI values obtained by varying the top 5 most-influential parameters and by varying all 39 parameters.[147] Our null hypothesis was that the sets of QOI values (shown graphically by the orange and black curves in **Fig. 6**) originated from the same underlying distribution. We obtained a very small KS test p-value (approximately $1.786 \times 10^{-181}$), which implies that the two sets of QOI values were likely drawn from different distributions. This indicates that we would need to vary more than 5 parameters to obtain a set of QOI values similar to that obtained by varying all 39 parameters. But since the KS test is a statistically consistent test, there is also an effect due to the large number (250,000) of samples we used to generate each set of QOIs, which will cause the KS test to even detect very small differences in the cumulative density functions (CDFs) of the orange and black curves.[148] To be precise, for large sets of values and for a given significance level $\alpha$, the null hypothesis is rejected if $D_{n,m} > c(\alpha)\sqrt{\frac{n+m}{n \cdot m}}$ where $D_{n,m}$ is the maximum vertical distance between the two CDFs, $c(\alpha)$ is a function of the significance level that increases as $\alpha$ decreases, and $n, m$ are the respective sizes of each sample.[148] Thus, when sample sizes increase, the quantity $c(\alpha)\sqrt{\frac{n+m}{n \cdot m}}$ decreases, and $D_{n,m}$ approaches the true maximum vertical distance between the two CDFs, leading to rejection of the null hypothesis. The smallest number of top-influential parameters required to obtain a p-value above 0.05 in our case was 18, which resulted in a p-value of approximately 0.201. The CDFs used for each of these two KS tests (varying 5 parameters vs. 39 parameters, and varying 18 parameters vs. 39 parameters) are shown in **Fig. S2** in the Supplementary Material.

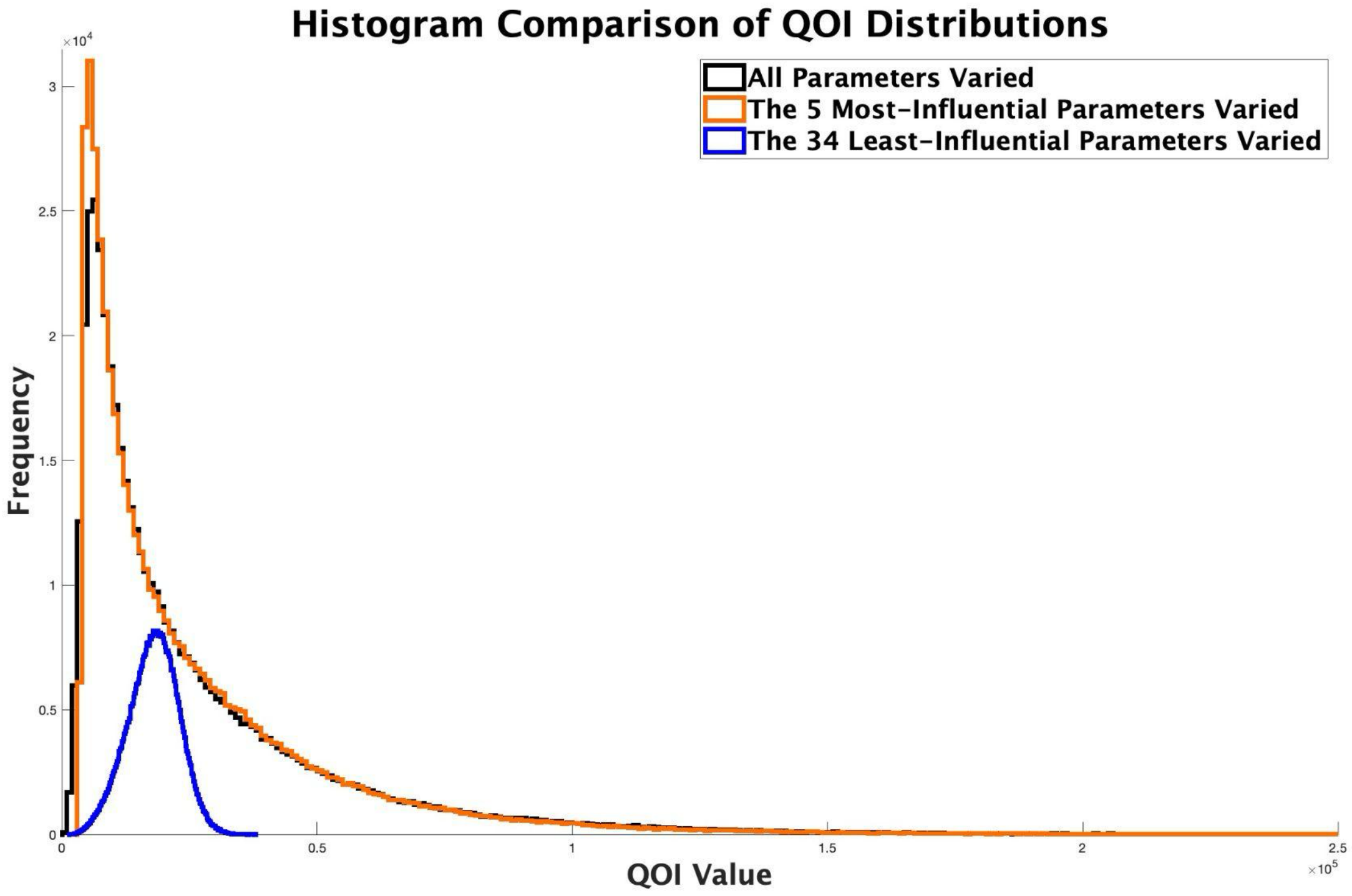

**Fig. 6. Histograms of QOI values when various parameters are allowed to vary.** To compare model behavior produced by all parameters versus model behavior produced by the top 5 most-influential parameters, we plotted QOI values (TNBC tumor size at Day 168) obtained by allowing all parameters to vary versus only allowing the top 5 most-influential parameters to vary. The black curve is a histogram that was obtained by randomly generating 250,000 parameter sets, where each of the 39 parameters was allowed to vary in the same ranges that were specified for the sensitivity analysis. That is, each parameter was given a range between 0.5 times its nominal value and 1.5 times its nominal value. The orange curve is a histogram of QOI values produced by randomly generating 250,000 parameter sets, where only the top 5 most-influential parameters were allowed to vary, while the 34 least-influential parameters remained fixed at their nominal values. For contrast, the blue curve is a histogram of QOI values produced by randomly generating 250,000 parameters sets, where only the 34 least-influential parameters were allowed to vary, and the top 5 were fixed at their nominal values. The variability in the QOI when the top 5 most-influential parameters were varied closely aligned with the QOI variability when all of the parameters were varied. This figure is truncated on the right; the full figure is available in the supplementary material (**Fig. S1)**. Calculations and plots were obtained using MATLAB 2024b.

The sensitivity analysis results are useful for indicating which parameters have more influence than others, but they do not elucidate specific relationships between certain parameters. To show these relationships, we plotted each possible pair of the top 5 most-influential parameters and colored each point in the plot by its QOI value (**Fig. 7**). Specifically, we took a subset of 10,000 parameter sets used for the global sensitivity analysis and the QOI values they produced, and plotted values of one parameter on the horizontal axis, and values of another parameter on the vertical axis, with each point colored by a heat map color to depict its QOI value. These projections onto two-dimensional planes in parameter space help us understand any patterns that pairs of parameters have on our QOI. For example, in **Fig. 7** the top left pairwise plot shows that even if a tumor has a relatively fast proliferation rate, a weak boosting effect from fibroblasts could potentially result in a smaller TNBC tumor. That is, CAF cytokine secretion represents a potential therapeutic target, even in the context of aggressive tumor growth. This analysis therefore provides insight into which biological interactions may offer the greatest leverage for combination treatment strategies, beyond ranking parameters in isolation.

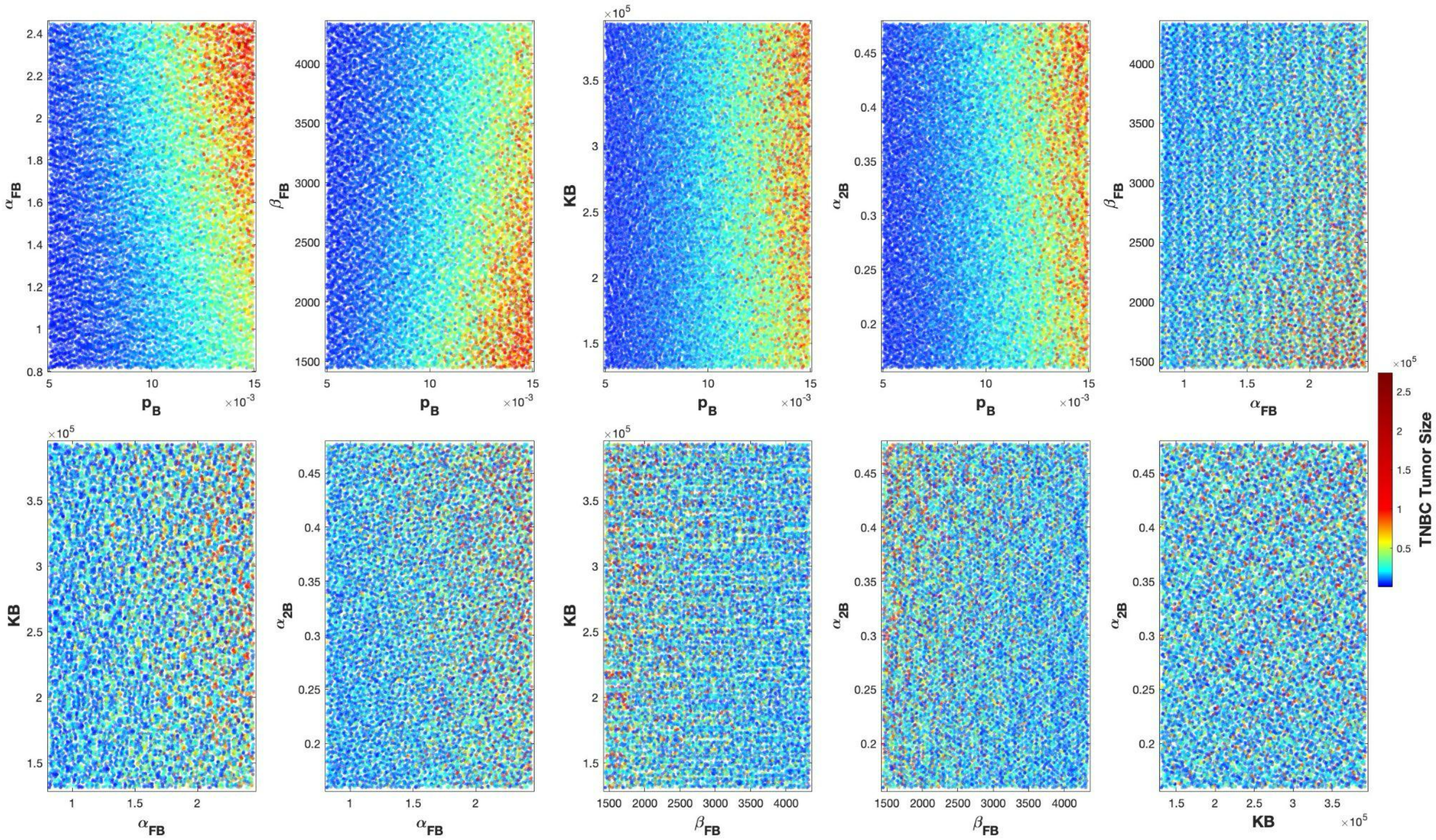

**Fig. 7. Pairwise relationships between the most-influential parameters.** Projection plots in two-dimensional subsets of parameter space, with each point colored by its corresponding QOI value (TNBC tumor size at Day 168), showing pairwise interactions between the top 5 most-influential parameters. Each plot shows 10,000 points, representing 10,000 parameter sets (generated using Sobol' sampling) and the corresponding QOI values they produced. On each plot, two influential parameters were selected for the axes, and each of the 10,000 points were positioned according to their values for those two parameters, and colored according to the size of the QOI those parameter values produced. As indicated by the colorbar, larger QOI values are represented by red, and smaller QOI values are represented by blue. Since only a few of the 10,000 points were very large, the color bar was set to color all QOI values above $1 \times 10^5$ a shade of red. This was done to enable easier interpretation of the patterns in each plot to be more-easily interpreted. The plots were created with MATLAB 2024b.

## Discussion

In this work, we developed a novel ODE model to describe key interactions within the TNBC TME. By modeling five representative cell populations and their primary interactions, we captured biologically-meaningful dynamics that align with current understanding of TNBC progression and immune evasion. After rigorous estimation of model parameters, global sensitivity analysis identified a small subset of parameters that account for the great majority of effects on tumor burden over a clinically-relevant timeframe. These parameters represent potential therapeutic targets that can be evaluated in future computational, experimental, or clinical studies.

Notably, highly-influential parameters are involved in CAF and M2 macrophage-mediated mechanisms, both of which are increasingly recognized as important contributors to TNBC aggressiveness,[69,149] and are being investigated as therapeutic targets in clinical studies.[110,150] The prominence of these mechanisms in the sensitivity analysis supports their relevance as promising areas for future investigation in combination therapy, rather than as monotherapies, since our analysis points to parameters with more influence on tumor size that are involved in separate mechanisms.

In the future, we can mathematically simulate these current or theoretical therapies using our model to predict expected outcomes for various regimens. We can also use our model and optimal control methods to optimize any combination drug regimens for triple-negative breast cancer patients. Our rigorous parameter estimations can also be used by other modelers who study different types of breast cancer, other cancers, and other diseases, supporting the wider scientific community.

The primary limitations of this work arise from parameter uncertainty, model size and uncertainty, and limited data availability. Comprehensive, quantitative data describing immune cell and stromal dynamics in human TNBC tumors remain limited, necessitating the use of data from related experimental systems to estimate certain parameters. Although these choices introduce uncertainty, the sensitivity analysis explores the behavior of the system with a range of different parameter values, and we focused on the relative influence of parameters rather than precise numerical predictions. Additionally, due to the heterogeneity of breast cancer tumors, including TNBC subtypes,[151] there is a need for data to describe immune cell and stromal cell densities to inform estimation of initial values.

Furthermore, reducing the TME to five interacting populations necessarily excludes other relevant cell types and signaling pathways. This simplification was required due to limited data availability, to ensure model identifiability and interpretability. It results in parameters that represent net effects in the system, which includes effects beyond their descriptions, due to the absence of other components that may also be relevant. Future extensions of this work may incorporate: 1) model calibration and validation with experimental data; 2) model refinement by adding cell populations and/or cell-cell interactions; or 3) simulation and optimization of treatment effects.

Code Availability: All code used to generate figures and tables with a detailed README file can be publicly accessed at https://github.com/KyleAdams26/TNBC.

Author Contributions:
KA reviewed the literature, developed the initial draft of the model, revised the model, estimated parameters, wrote the code, performed computations, and wrote and revised the manuscript. JB revised the model, estimated parameters, interpreted results, and wrote and revised the manuscript. SA reviewed the literature, revised figures, reviewed the code, and wrote and revised the manuscript. AB interpreted results, and wrote and revised the manuscript. MG revised the model, interpreted results, and revised the manuscript. HM conceived the study and the theoretical framework, supervised the project, reviewed the literature, revised the model, reviewed the code, and wrote and revised the manuscript.

## Supplemental Material

### Sensitivity Index Tables

**Table S1: Total Sensitivity Indices, $S_T$, with Different Numbers of Base Samples**

| Parameter | Indices from 175,000 Samples | Indices from 150,000 Samples | Indices from 125,000 Samples | Indices from 100,000 Samples |
|---|---|---|---|---|
| $p_B$ | 0.7498169832 | 0.7504181573 | 0.75055292 | 0.7478398025 |
| $\alpha_{FB}$ | 0.2039521835 | 0.2039979498 | 0.2027910049 | 0.2021303876 |
| $\beta_{FB}$ | 0.1348050234 | 0.1348737305 | 0.1344004911 | 0.1348904215 |
| $K_B$ | 0.03185837667 | 0.03184826609 | 0.03143385635 | 0.03212701711 |
| $\alpha_{2B}$ | 0.02798995512 | 0.0278140573 | 0.02597266152 | 0.02505309046 |
| $\delta_B$ | 0.0208431921 | 0.0209617052 | 0.02089986807 | 0.02133431191 |
| $s_F$ | 0.01974118216 | 0.0192915876 | 0.01911589037 | 0.01860060498 |
| $\alpha_{CB}$ | 0.01312429316 | 0.01312146974 | 0.01335880712 | 0.01303590778 |
| $\alpha_{B2}$ | 0.01172636239 | 0.01170430464 | 0.01156540689 | 0.01164802985 |
| $p_C$ | 0.01027792839 | 0.01012421801 | 0.01008850233 | 0.0101089608 |
| $\beta_{2B}$ | 0.009636705451 | 0.009476314285 | 0.009456533683 | 0.009474608904 |
| $\delta_2$ | 0.008833135646 | 0.008739061912 | 0.008747747602 | 0.008692556996 |
| $\delta_C$ | 0.007646046507 | 0.007714760697 | 0.008089638402 | 0.008227199305 |
| $\beta_{CB}$ | 0.006535875212 | 0.006501564957 | 0.006630131729 | 0.006565198786 |
| $K_C$ | 0.006315037256 | 0.006199518298 | 0.006273608007 | 0.006181008496 |
| $\beta_{B2}$ | 0.002701998922 | 0.00278175296 | 0.002822725127 | 0.0028975856 |
| $\alpha_{BCB}$ | 0.001711954325 | 0.001831112852 | 0.001841970565 | 0.001567100852 |
| $\alpha_{2CB}$ | 0.001243609115 | 0.001229541266 | 0.00131318775 | 0.001270893036 |
| $\delta_F$ | 0.0009651154292 | 0.0009312629996 | 0.001002286496 | 0.001009026659 |
| $\beta_{BCB}$ | 0.0007277059719 | 0.0007472947576 | 0.000769684369 | 0.0007425119626 |
| $\delta_R$ | 0.000727632913 | 0.0006745094434 | 0.0006653330919 | 0.0006479255327 |
| $s_R$ | 0.0007207159842 | 0.0006790251068 | 0.000643797839 | 0.0006430765229 |
| $\alpha_{F2}$ | 0.0006939536616 | 0.000614242199 | 0.0006117718675 | 0.0006854699535 |
| $\beta_{2CB}$ | 0.0005376135743 | 0.0005078338345 | 0.0005675383647 | 0.0004816456045 |
| $\alpha_{RCB}$ | 0.000527822781 | 0.0004927596162 | 0.0003832765435 | 0.0003513980378 |
| $\beta_{F2}$ | 0.0004746610803 | 0.0004844540207 | 0.0004593671522 | 0.0004279661751 |
| $\beta_{RC}$ | 0.0004151404953 | 0.0003124321904 | 0.000328839373 | 0.0002679983866 |
| $\alpha_{RC}$ | 0.0003757060878 | 0.0003739473435 | 0.0002880505071 | 0.0003094205999 |
| $\beta_{RCB}$ | 0.0002984180542 | 0.0003065551984 | 0.0002808446297 | 0.0002279437243 |
| $\alpha_{BR}$ | $6.87 * 10^{-5}$ | $6.88 * 10^{-5}$ | $7.63 * 10^{-5}$ | $9.59 * 10^{-5}$ |
| $\gamma_2$ | $2.96 * 10^{-5}$ | $-1.67 * 10^{-5}$ | $1.43 * 10^{-5}$ | $4.35 * 10^{-5}$ |

| $\beta_{BR}$ | 2.52 * $10^{-5}$ | 9.55 * $10^{-7}$ | 3.04 * $10^{-6}$ | 1.40 * $10^{-5}$ |
|---|---|---|---|---|
| $\alpha_{2C}$ | 1.00 * $10^{-5}$ | 1.25 * $10^{-5}$ | 5.26 * $10^{-6}$ | -9.19 * $10^{-7}$ |
| $\alpha_{C2}$ | 5.28 * $10^{-6}$ | 1.80 * $10^{-6}$ | 3.67 * $10^{-5}$ | 3.90 * $10^{-5}$ |
| $\beta_{2R}$ | 3.68 * $10^{-6}$ | 5.57 * $10^{-7}$ | 2.94 * $10^{-6}$ | 4.12 * $10^{-6}$ |
| $K_F$ | -3.92 * $10^{-6}$ | 1.95 * $10^{-6}$ | 1.01 * $10^{-5}$ | -1.82 * $10^{-5}$ |
| $\beta_{C2}$ | -1.13 * $10^{-5}$ | -1.82 * $10^{-6}$ | -2.19 * $10^{-5}$ | -1.08 * $10^{-5}$ |
| $\beta_{2C}$ | -1.73 * $10^{-5}$ | -1.32 * $10^{-5}$ | -1.52 * $10^{-5}$ | -1.27 * $10^{-5}$ |
| $\alpha_{2R}$ | -2.06 * $10^{-5}$ | -1.93 * $10^{-5}$ | -2.02 * $10^{-5}$ | -7.92 * $10^{-6}$ |

**Table S1.** This table shows the total Sobol' sensitivity index value for each parameter, for different numbers of base samples. The table is in order of descending total Sobol' sensitivity index calculated from 175,000 base samples, which appear in the left-most column of index values. We began our Sobol' sensitivity analyses with 100,000 base samples (right-most column of this table), and increased them in increments of 25,000 (from the right-most column to the left-most column) to study the ordering of top parameters. The 20 most-influential parameters do not change ordering between 100,000 and higher numbers of base samples.

| **Table S2: First-Order Sensitivity Indices, $S_1$, with Different Numbers of Base Samples** | | | | |
|---|---|---|---|---|
| **Parameter** | **Indices from 175,000 Samples** | **Indices from 150,000 Samples** | **Indices from 125,000 Samples** | **Indices from 100,000 Samples** |
| $p_B$ | 0.5545189526 | 0.5546130067 | 0.5556180915 | 0.5546274206 |
| $\alpha_{FB}$ | 0.08226329078 | 0.08236131074 | 0.08263484018 | 0.08225549766 |
| $\beta_{FB}$ | 0.05379395767 | 0.05386586762 | 0.05490307768 | 0.05510605962 |
| $K_B$ | 0.005498107267 | 0.005239603847 | 0.005678324988 | 0.005887669885 |
| $\alpha_{2B}$ | 0.009305282795 | 0.009316847618 | 0.00922086923 | 0.009118477024 |
| $\delta_B$ | 0.011426453 | 0.011246594 | 0.01125410313 | 0.01131909745 |
| $s_F$ | 0.006310363952 | 0.006305880664 | 0.006259546122 | 0.006281230161 |
| $\alpha_{CB}$ | 0.006781484283 | 0.006925938853 | 0.006743246465 | 0.006399389308 |
| $\alpha_{B2}$ | 0.004847161015 | 0.004978468331 | 0.004917385927 | 0.004944333789 |
| $p_C$ | 0.004721060198 | 0.004752330315 | 0.004738120072 | 0.004750027388 |
| $\beta_{2B}$ | 0.003626868735 | 0.003665980006 | 0.003670006174 | 0.00372432587 |
| $\delta_2$ | 0.003594130799 | 0.003513306095 | 0.003555263805 | 0.003608005984 |
| $\delta_C$ | 0.003323538508 | 0.003267475622 | 0.003120227178 | 0.003190444718 |
| $\beta_{CB}$ | 0.003171488443 | 0.003072252077 | 0.003099690476 | 0.003029736551 |
| $K_C$ | 0.003162788623 | 0.003230000732 | 0.003212802181 | 0.003206866017 |
| $\beta_{B2}$ | 0.001633377991 | 0.001651480026 | 0.001656203324 | 0.001744369402 |

| $\alpha_{BCB}$ | 0.0009002588181 | 0.0009134694738 | 0.0009634724292 | 0.0008939061308 |
|---|---|---|---|---|
| $\alpha_{2CB}$ | 0.0004282369505 | 0.0004196424333 | 0.0004042223747 | 0.0003502220683 |
| $\delta_F$ | 0.000407177094 | 0.000411393942 | 0.0004209488827 | 0.0003998549984 |
| $\beta_{BCB}$ | 0.000314184621 | 0.0003281594597 | 0.000312899262 | 0.000354707205 |
| $\delta_R$ | 0.0002248140664 | 0.0002655640473 | 0.0002457963075 | 0.0002002964216 |
| $s_R$ | 0.0003088832266 | 0.0003211742405 | 0.0002764140069 | 0.0002852262825 |
| $\alpha_{F2}$ | 0.0002612020286 | 0.0002594006339 | 0.0002257028935 | 0.0002267605612 |
| $\beta_{2CB}$ | 0.0002367831803 | 0.0002192882914 | 0.0001875812995 | 0.0001400161653 |
| $\alpha_{RCB}$ | 0.0005055865845 | 0.0006497612585 | 0.0007790794844 | 0.0007908734433 |
| $\beta_{F2}$ | 0.0001783781772 | 0.0001815764784 | 0.0001752553398 | 0.0002060536844 |
| $\beta_{RC}$ | 0.0001189797278 | 0.0001205120526 | 0.0001351530569 | 0.0001617552024 |
| $\alpha_{RC}$ | 0.0002099515903 | 0.0002100442471 | 0.0002130970171 | 0.0002234092083 |
| $\beta_{RCB}$ | 0.0001128103766 | 0.0001114844355 | 0.0001309158439 | 0.0001575735381 |
| $\alpha_{BR}$ | $2.62 * 10^{-5}$ | $1.85 * 10^{-5}$ | $5.78 * 10^{-6}$ | $2.52 * 10^{-5}$ |
| $\gamma_2$ | $1.93 * 10^{-5}$ | $2.40 * 10^{-5}$ | $1.30 * 10^{-5}$ | $3.82 * 10^{-6}$ |
| $\beta_{BR}$ | $2.78 * 10^{-5}$ | $2.88 * 10^{-5}$ | $3.48 * 10^{-5}$ | $3.99 * 10^{-5}$ |
| $\alpha_{2C}$ | $-3.89 * 10^{-6}$ | $-5.63 * 10^{-7}$ | $1.26 * 10^{-7}$ | $9.86 * 10^{-6}$ |
| $\alpha_{C2}$ | $2.50 * 10^{-5}$ | $2.34 * 10^{-5}$ | $2.08 * 10^{-5}$ | $2.15 * 10^{-5}$ |
| $\beta_{2R}$ | $2.02 * 10^{-6}$ | $-4.65 * 10^{-7}$ | $2.02 * 10^{-6}$ | $-7.37 * 10^{7}$ |
| $K_F$ | $2.29 * 10^{-6}$ | $3.65 * 10^{-6}$ | $-3.80 * 10^{-7}$ | $5.88 * 10^{-6}$ |
| $\beta_{C2}$ | $2.25 * 10^{-5}$ | $2.21 * 10^{-5}$ | $1.99 * 10^{-5}$ | $2.55 * 10^{-5}$ |
| $\beta_{2C}$ | $-2.01 * 10^{-6}$ | $1.67 * 10^{-6}$ | $-1.75 * 10^{-6}$ | $3.49 * 10^{-6}$ |
| $\alpha_{2R}$ | $9.41 * 10^{-6}$ | $1.13 * 10^{-5}$ | $1.41 * 10^{-5}$ | $1.13 * 10^{-5}$ |

**Table S2.** This table shows the first-order Sobol' sensitivity index value for each parameter, for different numbers of base samples. The table is in order of descending *total* Sobol' sensitivity index values calculated from 175,000 base samples, which appear in the left-most column of index values in Table ST. We began our Sobol' sensitivity analyses with 100,000 base samples, and increased them in increments of 25,000 to study the ordering of top parameters. The 20 most-influential parameters do not change ordering between 100,000 and higher numbers of base samples.

Comparing QOI Distributions

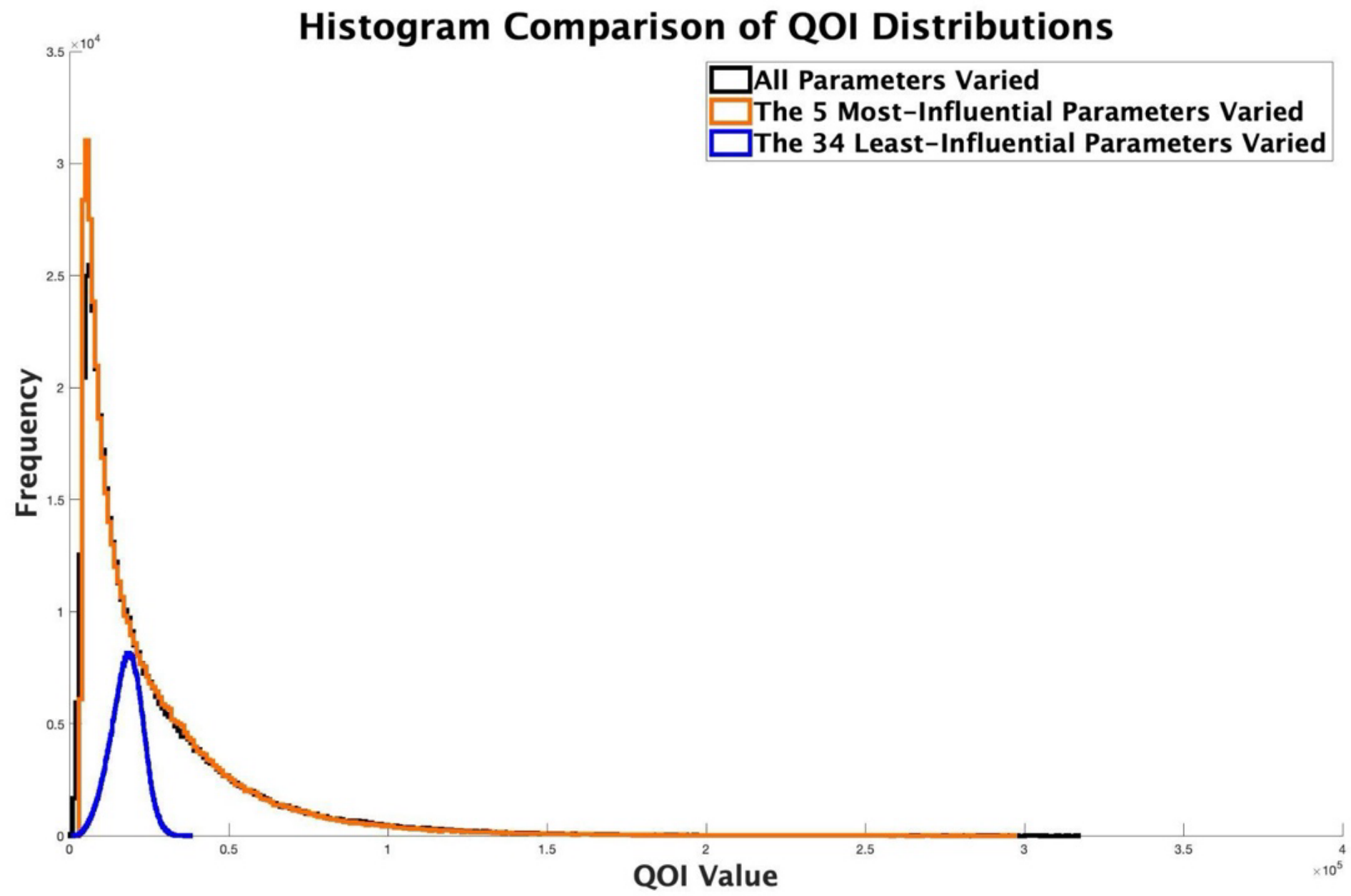

**Fig. S1.** Histogram comparison of QOI distributions with different parameters varied, with all values included in the plot (no truncation of axes).

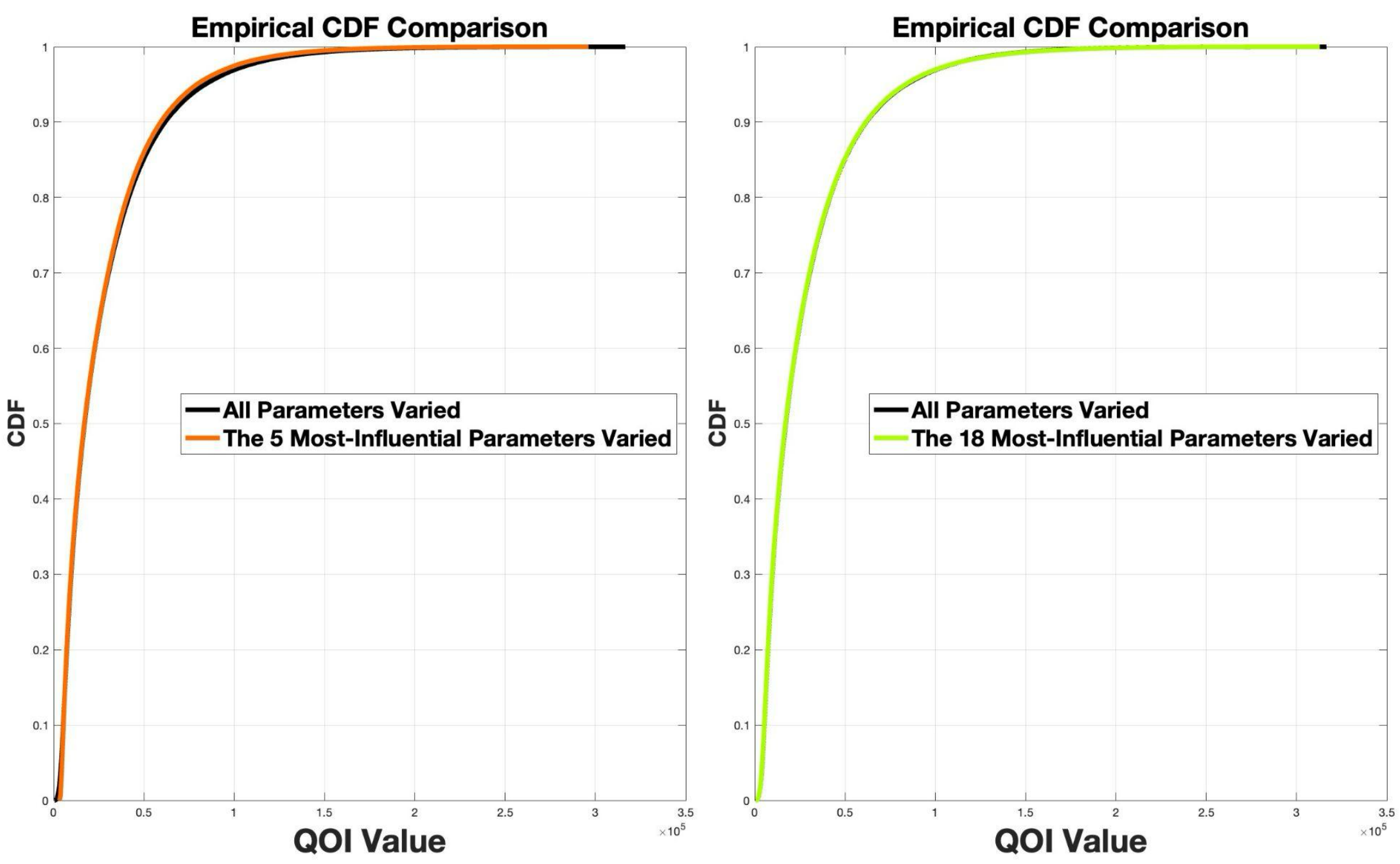

**Fig. S2.** CDF comparisons of QOI distributions. The p-value obtained from a two-sample KS test comparing the CDFs of the black curve and the orange curve was approximately $1.786 \times 10^{-181}$. The p-value obtained from a two-sample KS test comparing the CDFs of the black curve and the green curve was approximately 0.201.